\documentclass[journal,final,twocolumn,10pt,twoside]{IEEEtranTMBMC}
\normalsize
\IEEEoverridecommandlockouts
\usepackage{cite}
\usepackage{amsmath,amssymb,amsfonts}
\usepackage{algorithmic}
\usepackage{graphicx}
\usepackage{textcomp}
\usepackage{xcolor}
\usepackage{cases}
\usepackage{orcidlink}

\DeclareMathOperator*{\argmax}{arg\,max}

\begin{document}
\bstctlcite{IEEEexample:BSTcontrol} 

\title{A Computational Approach for the Characterization of Airborne Pathogen Transmission in Turbulent Molecular Communication Channels \\ 
\thanks{This work was presented in part at the 2022 IEEE Global Communications Conference (GLOBECOM), Rio de Janeiro, Brazil \cite{gulec2022characterization}.}%
\thanks{The work of Fatih Gulec was supported in part by German Academic Exchange Service (DAAD) and in part by a Discovery grant from the Natural Sciences and Engineering Research Council of Canada. The work of Falko Dressler was supported by the project MAMOKO funded by the German Federal Ministry of Education and Research (BMBF) under Grant 16KIS0917. Andrew W. Eckford was funded in part by a Discovery grant from the Natural Sciences and Engineering Research Council of Canada.
}}


\author{Fatih~Gulec$^{\orcidlink{0000-0003-1031-6108}}$,~\IEEEmembership{Member,~IEEE,}
	Falko~Dressler,~\IEEEmembership{Fellow,~IEEE,}
	and~Andrew~W.~Eckford,~\IEEEmembership{Senior~Member,~IEEE}
	\thanks{Fatih Gulec is with the Department of Electrical Engineering and Computer Science, York University, Toronto, Ontario, Canada and with the School of Electrical Engineering and Computer Science, TU Berlin, Berlin, Germany, e-mail: (fgulec@yorku.ca).
		
	Falko Dressler is with the School of Electrical Engineering and Computer Science, TU Berlin, Berlin, Germany, e-mail: (dressler@ccs-labs.org).
	
	Andrew W. Eckford is with the Department of Electrical Engineering and Computer Science, York University, Toronto, Ontario, Canada e-mail: (aeckford@yorku.ca).		

}
}

\maketitle

\begin{abstract}
Airborne pathogen transmission mechanisms play a key role in the spread of infectious diseases such as COVID-19. In this work, we propose a computational fluid dynamics (CFD) approach to model and statistically characterize airborne pathogen transmission via pathogen-laden particles in turbulent channels from a molecular communication viewpoint. To this end, turbulent flows induced by coughing and the turbulent dispersion of droplets and aerosols are modeled by using the Reynolds-averaged Navier-Stokes equations coupled with the realizable $k-\epsilon$ model and the discrete random walk model, respectively. Via simulations realized by a CFD simulator, statistical data for the number of received particles are obtained. These data are post-processed to obtain the statistical characterization of the turbulent effect in the reception and to derive the probability of infection. Our results reveal that the turbulence has an irregular effect on the probability of infection, which shows itself by the multi-modal distribution as a weighted sum of normal and Weibull distributions. Furthermore, it is shown that the turbulent MC channel is characterized via multi-modal, i.e., sum of weighted normal distributions, or stable distributions, depending on the air velocity.
\end{abstract}


\section{Introduction}
Airborne transmission is an important contagion mechanism of pathogens (e.g., viruses, bacteria) in the spread of infectious diseases such as influenza and COVID-19 \cite{ seminara2020biological, zhao2022airborne, ai2018airborne}. In airborne transmission, infectious diseases spread by pathogen-laden particles (droplets and aerosols) through respiratory activities such as breathing, coughing, sneezing and speaking \cite{mittal2020flow, balachandar2020host, abkarian2020speech}. Especially, coughing and sneezing induce turbulent flows due to the high initial emission velocity of particles from the mouth \cite{bhagat2020effects}. Although there are analytical methods in the literature of fluid dynamics, these models simplify the effects of turbulence \cite{bourouiba2014violent, de2021evolution}. Instead, computational fluid dynamics (CFD) simulators are widely employed to model  airborne transmission via droplets in order to model turbulent flows and airflow-particle interactions more realistically with a cost of computational load. In \cite{pendar2020numerical, vuorinen2020modelling,liu2021investigation}, coughing and the dispersion of droplets are modeled by using the large eddy simulation (LES) turbulence model. In \cite{busco2020sneezing} and \cite{dbouk2020coughing}, flows as a result of sneezing/coughing are modeled with the Reynolds-averaged Navier-Stokes (RANS) equations coupled with the realizable $k-\epsilon$ and $k-\omega$ turbulence models, respectively. Studies in \cite{dbouk2021airborne, chea2021assessment, li2020dispersion} also employ simulations using the RANS equations for respiratory releases. In addition, the stochastic turbulent dispersion of droplets emitted by a cough are evaluated for an indoor scenario by using the LES turbulence model in \cite{trivedi2021estimates}.

The research in fluid dynamics literature focuses on the propagation of droplets after the emission but not the reception elaborately. However, there is a similarity between air-based molecular communication (MC) systems and airborne transmission of pathogens  \cite{khalid2019communication, gulec2021fluid, gulec2020droplet, schurwanz2021duality,bhattacharjee2022digital}. This analogy can be considered as a MC problem to detect the airborne viruses with biological sensors \cite{khalid2019communication, khalid2020modeling,amin2021viral,chen2022detection}. In addition, the transfer of pathogen-laden particles between humans can be considered as a way of communication. Hence, the usage of MC is proposed to model the airborne transmission with a holistic approach \cite{gulec2021molecular, schurwanz2021duality, schurwanz2021infectious, barros2021molecular, koca2021molecular, pal2021vivid, lotter2021statistical, felicetti2021molecular,gulec2022mobile}.

The studies in \cite{schurwanz2021duality} and \cite{schurwanz2021infectious} lay the theoretical and experimental foundations of dualities between pathogen-laden droplet propagation and MC. In \cite{koca2021molecular} and \cite{pal2021vivid}, the transmission mechanism of Severe Acute Respiratory Syndrome-Corona Virus-2 in the respiratory system is modeled with a MC perspective. In \cite{lotter2021statistical}, a statistical model for the spread of viruses through imperfectly fitted masks is proposed. In \cite{gulec2021molecular}, an end-to-end MC  system model which considers the coughed droplets as a cloud with a probabilistic approach is proposed to model the airborne transmission between two humans. This approach also enables an analytical derivation of infection probability which can be used in transmission and epidemiology models. In \cite{gulec2022mobile}, the analogy between human groups and telecommunication networks is employed for the mobile human ad hoc network architecture to estimate the time course of epidemics by exploiting MC. However, none of the aforementioned works employs the turbulent flows and aerosols together with a MC perspective. Furthermore, the statistical characterization of received particles under turbulent flows and the corresponding analytical expression for the probability of infection is not known.

In this paper, a CFD approach is proposed to model the airborne transmission of cough droplets and aerosols together with turbulent flows between an infectious and susceptible human which are the transmitter (TX) and receiver (RX), respectively. In this approach, the channel is modeled as a turbulent two-phase flow medium which comprises the movement of airflows with turbulence (continuous phase) and the motions of particles (discrete phase) interacting with these airflows and RX. The turbulence is modeled by employing the RANS equations coupled with the realizable $k-\epsilon$ model and the turbulent dispersion of particles is tracked in the CFD simulator by using the discrete random walk model.

In order to observe and characterize the effect of turbulence in airborne transmission, extensive CFD simulations are executed. Thus, statistical data for the received particles are obtained and employed to derive the probability of infection. Our statistical analysis shows that modeling the effect of turbulence on infections is not straightforward, since the probability density function (pdf) of received particles are modeled by multi-modal distributions, i.e., weighted addition of normal and Weibull distributions. Additionally, it is shown that aerosols are more effective than large droplets for $1.5$ m in reception and the increment of the ambient airflow increases the probability of infection. Moreover, the end-to-end impulse response of the turbulent air-based MC channel is characterized. While for lower air velocities ($v_{air}=0.1$ m/s and $v_{air}=0.3$ m/s) its pdf is multi-modal, i.e., weighted sum of normal distributions, it has a uni-modal stable distribution for $v_{air}=0.5$ m/s.

Our main contributions can be summarized as follows:
\begin{itemize}
	\item A CFD-based MC approach is proposed to realistically model the effect of turbulent flows induced by coughing in airborne infectious disease transmission.
	\item The pdf of received number of particles (both droplets and aerosols) is characterized for an indoor airborne transmission scenario via coughing and the corresponding analytical expression for the probability of infection is derived. The underlying multi-modal distributions of these pdfs are revealed for different ambient air velocities.
	\item It is shown that aerosols are more important than large droplets at $1.5$ m distance for infection.
	\item The air-based turbulent MC channel is characterized. It is shown that the end-to-end system response can have multi-modal or uni-modal distributions depending on the ambient air velocity.
\end{itemize}

The rest of the paper is organized as follows. In Section \ref{SM}, the end-to-end system model including turbulent flow, particle tracking and receiver is detailed. Section \ref{CFD_Setup} provides the details about the employed CFD simulator setup. In Section \ref{SR}, the CFD simulation results are given. Then, the statistical characterization of the infection probability based on simulation results is elaborated in Section \ref{PoI}. In Section \ref{EtEIRC}, the end-to-end impulse response of the turbulent MC system is characterized and the paper is concluded in Section \ref{Conc}.

\section{System Model} \label{SM}
\begin{figure}[hb]
		\vspace{0.04in}
	\centering
	\includegraphics[width=0.8\columnwidth]{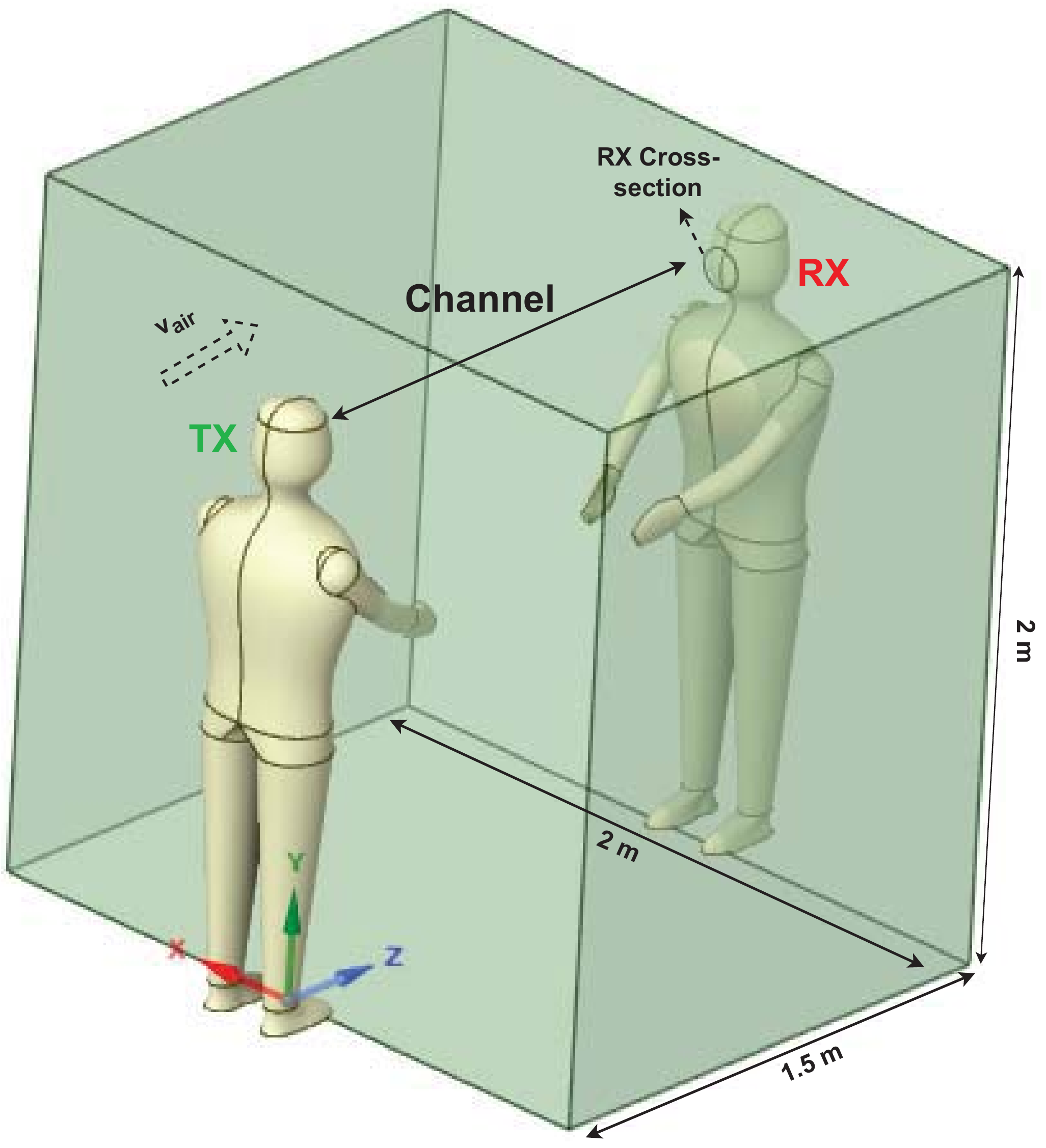}
	\caption{Airborne pathogen transmission scenario between two humans.}
	\vspace{-.5em}
	\label{Geo}
\end{figure}

\begin{figure}[!b]
	\centering
	\includegraphics[width=\columnwidth]{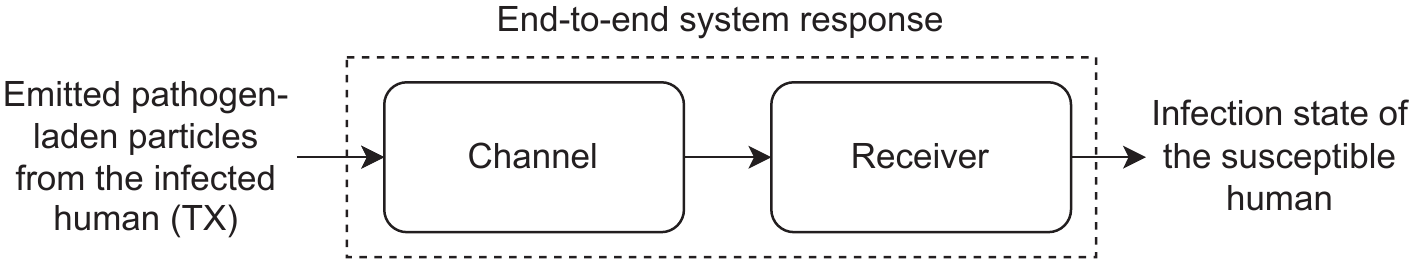}
	\caption{End-to-end system model.}
	\label{EtE}
	\vspace{-.5em}
\end{figure}

In this section, the 3-D system model for a scenario where the TX emits pathogen-laden spherical particles by coughing towards the RX in a room with an airflow as illustrated in Fig. \ref{Geo} is detailed. As shown in Fig. \ref{EtE}, emitted particles from the TX is considered as an impulsive input signal and propagate through the channel and sensed by the RX. Hence, the end-to-end system response is given as the output of the system, i.e., the infection state of the susceptible human. The end-to-end system model includes the details about the receiver model, turbulent two-phase channel model and emitted particle characterization which is given as follows.


\subsection{Emitted Pathogen-Laden Particle Size Distribution} \label{Particle_Distr}
\begin{figure}[!t]
	\centering
	\includegraphics[width=0.9\columnwidth]{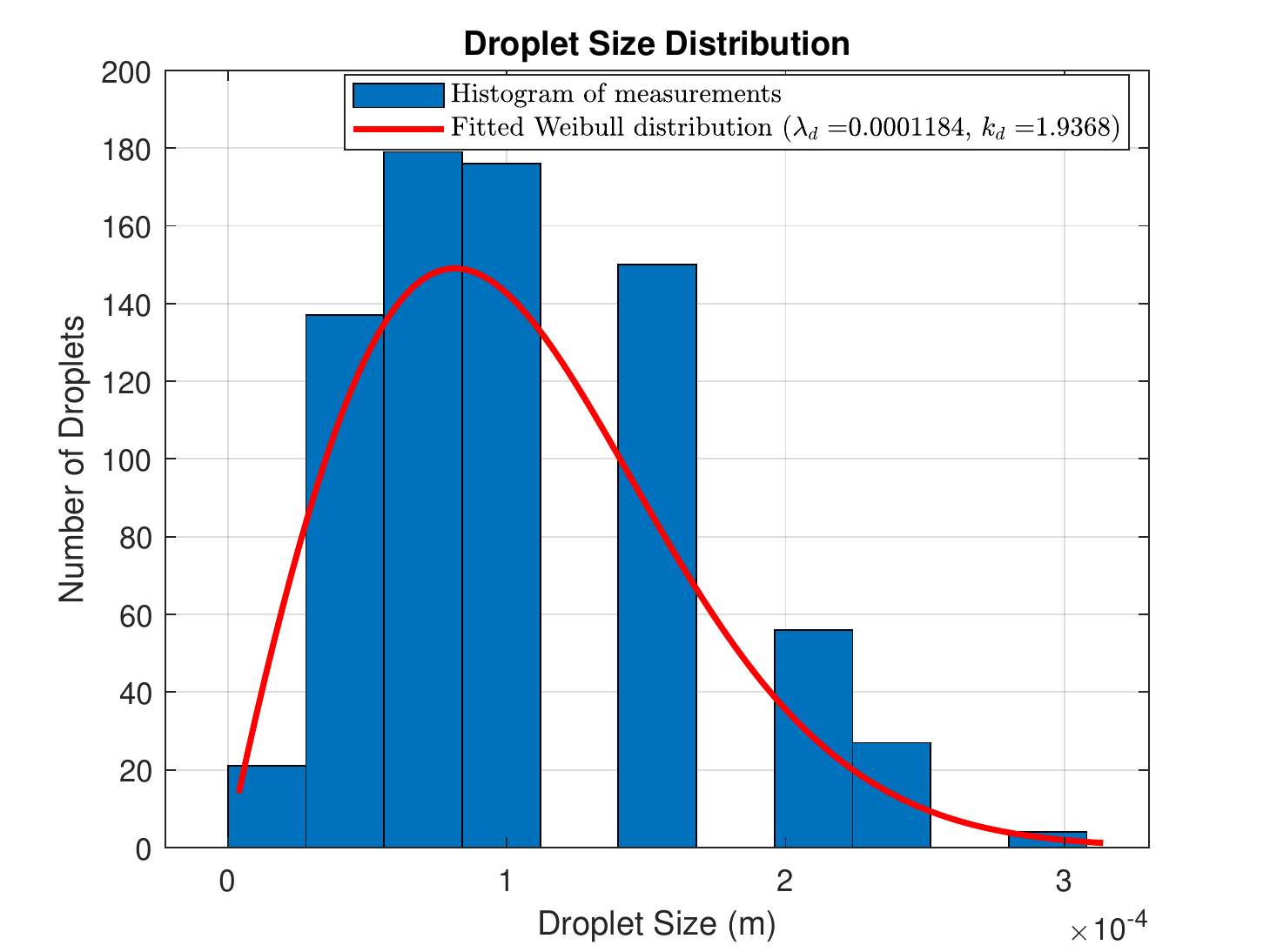}
	\caption{Histogram and fitted Weibull droplet size distribution according to \cite{xie2009exhaled}. $\lambda_d = 0.0001184$ with the estimation variance $5.553 \mathrm{e}-12$, $k_d = 1.9368$ with the estimation variance $0.00293$.}
	\label{droplet}
\end{figure}

\begin{figure}[!btph]
	\centering
	\includegraphics[width=0.9\columnwidth]{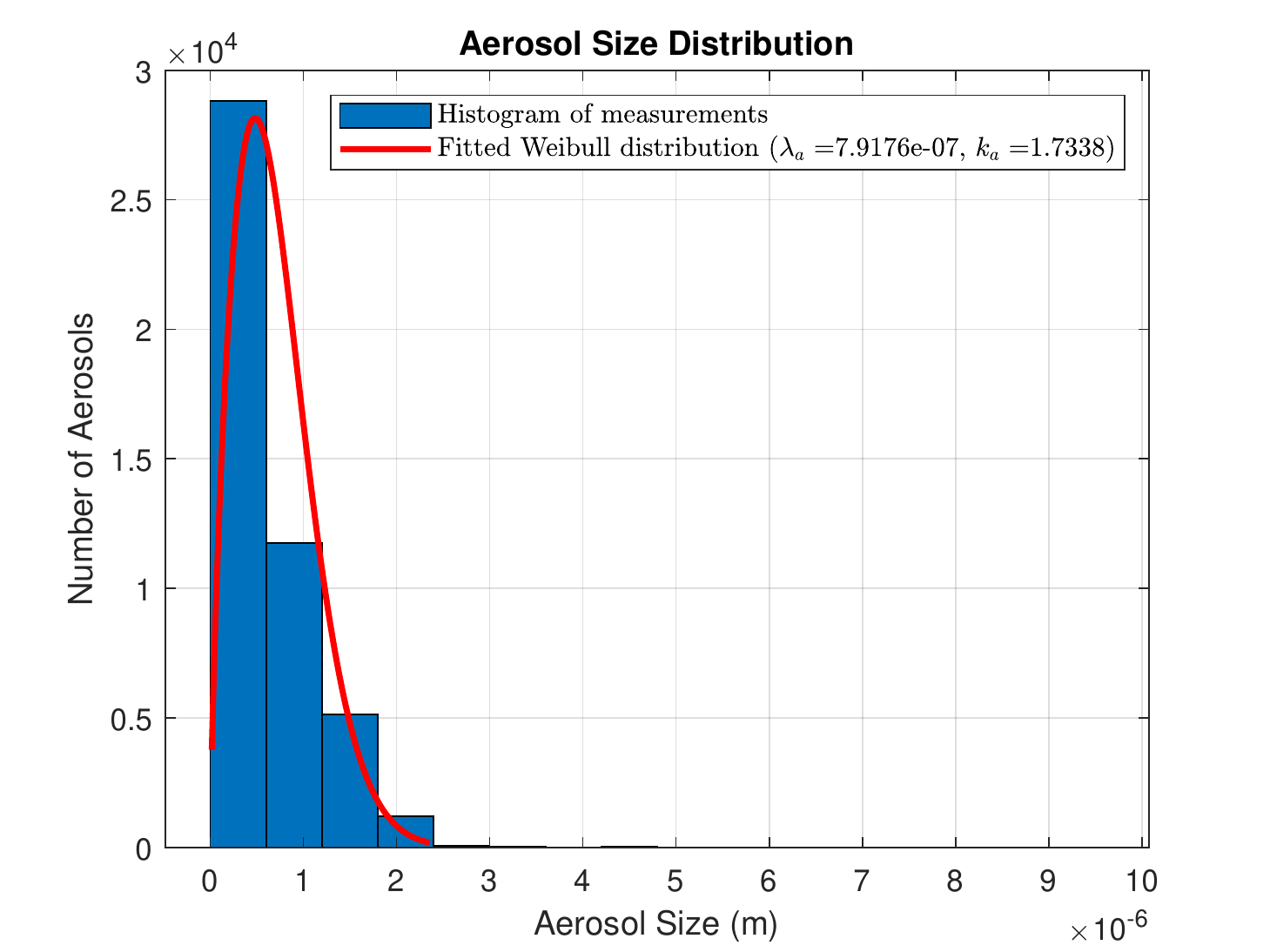}
	\caption{Histogram and fitted Weibull aerosol size distribution according to \cite{lindsley2012quantity}. $\lambda_a = 7.92\mathrm{e}-7$ with the estimation variance $5.0147\mathrm{e}-18$, $k_a = 1.7338$ with the estimation variance $2.785\mathrm{e}-5$.}
	\label{aerosol}
\end{figure}

In the literature of airborne transmission, only the propagation of large droplets ($\geq10$ $\mu$m) is taken into account. However, there is also evidence for the airborne transmission cases with aerosols ($<10$ $\mu$m) \cite{seminara2020biological}. Therefore, aerosols are also included to the emitted cough particles in our scenario. Experimental data in \cite{xie2009exhaled}, \cite{lindsley2012quantity} are used for the number and size (diameter) of spherical droplets and aerosols, respectively. These data are fitted by using the maximum likelihood estimation for a better implementation in the CFD simulator according to a Weibull distribution which has the probability density function (pdf) for data samples ($x \geq 0$) as given by
\begin{equation}
	f(x) = \frac{k}{\lambda} \left( \frac{x}{\lambda}\right)^{k-1} \textrm{e}^{-\left(\frac{x}{\lambda} \right)^k }, 
\end{equation}
where $k$ and $\lambda$ are shape and scale parameters, respectively. These parameters are estimated by using the maximum likelihood estimation which is based on finding the unknown parameter values ($\mathbf{y} = [y_1, y_2,...,y_n]^T$) which maximizes the log likelihood function ($ \mathcal{L}_n (\mathbf{y}; \mathbf{x}) $) as given by \cite{papoulis2002probability}
\begin{equation}
	\mathcal{L}_n (\mathbf{y};\mathbf{x}) = \log \left[f_n(\mathbf{x};\mathbf{y})\right] = \sum_{k=1}^{n} \log \left[f_k(x_k;\mathbf{y})\right],
\end{equation}
where $ \mathbf{x} $ represents the observed data samples, $\log(.)$ is the natural logarithm, and $ f_n(\mathbf{x};\mathbf{y}) $ is the joint pdf where data samples are assumed as independent and identically distributed random variables. Hence, the estimated parameters are determined according to the rule as given below:
\begin{equation}
	\hat{\mathbf{y}} = \argmax_{\mathbf{y} \in \Theta} \mathcal{L}_n (\mathbf{y};\mathbf{x})
\end{equation}
where $  \Theta $ is the parameter space. By applying this rule, the parameters are estimated in a least-square sense iteratively, i.e., the best fit for the estimated function is found by minimizing the sum of the squared error between the estimated values and actual values. Thus, the shape and scale parameters for the pdfs of droplet ($k_d, \lambda_d$) and aerosol sizes ($k_a, \lambda_a$) are estimated as $\lambda_d = 0.0001184$ with the estimation variance $5.553 \mathrm{e}-12$, $k_d = 1.9368$ with the estimation variance $0.00293$ as shown in Fig. \ref{droplet}, and $\lambda_a = 7.92\mathrm{e}-7$ with the estimation variance $5.0147\mathrm{e}-18$, $k_a = 1.7338$ with the estimation variance $2.785\mathrm{e}-5$ as shown in Fig. \ref{aerosol}, respectively. 


\subsection{Turbulent Two-Phase Flow Channel Model}
Pathogen-laden particles, which consist of large droplets and aerosols, are subject to some interactions with the air after the emission with an initial velocity from the TX. These interactions lead to turbulent airflows in the vicinity of the TX  in addition to the constant air velocity ($v_{air}$). The motion of particles rely on these airflows as well as other factors such as gravity and air drag. All of these motions in the MC channel can be examined as continuous (or gas) phase for airflows and discrete (or liquid) phase for particle movements.
\subsubsection{Continuous Phase}
Turbulence is considered by using the 3-D Navier-Stokes equations that determine airflow velocity components $u_i$ where $ i = {1,2,3} $ for $x$, $y$ and $z$ in Cartesian coordinates, respectively. To reduce the computational complexity for the solution of the Navier-Stokes equations, the RANS equations, which average the Navier-Stokes equations in time, are employed \cite{cebeci2005computational}. Instantaneous flow velocities are considered as the addition of the average values ($\overline{u_i}$) and the fluctuation values ($u_i'$), i.e., $u_i = \overline{u_i}+u_i'$. The RANS equations in tensor form for the average velocities are given by \cite{bi2021numerical}
\begin{equation}
		\frac{\partial \rho}{\partial t} + \frac{\partial}{\partial x_i} (\rho \overline{u_i})  =  0  
\label{RANS1}
\end{equation} \vspace{-0.5cm}
\begin{multline} 
		\frac{\partial}{\partial t} (\rho \overline{u_i}) + \frac{\partial}{\partial x_j} (\rho \overline{u_i} \overline{u_j}) = -\frac{\partial p}{\partial x_i}   +  \frac{\partial}{\partial x_j} \left[ \mu \left( \frac{\partial \overline{u_i}}{\partial x_j} + \frac{\partial \overline{u_j}}{\partial x_i} \right) \right. \\   \hspace{-1cm} \left. - \frac{2}{3}\delta_{ij} \frac{\partial \overline{u_k}}{\partial x_k}  \right] + \dfrac{\partial \left(-\rho \overline{u_i'u_j'} \right)}{\partial x_j},
\label{RANS2}
\end{multline}
where $\rho$ is the fluid density, $\mu$ is the dynamic viscosity, $p$ is the fluid pressure,  and $\delta_{ij}$ is the Kronecker delta function and $x_1$, $x_2$, and $x_3$ represent the Cartesian coordinates  $x$, $y$ and $z$, respectively. Please note that (\ref{RANS1})-(\ref{k-ep2}) are given in tensor (or Einstein) notation in order to write the equations in a shorter form. In addition, $\rho$ is assumed as a constant value whereas $p$ is a variable depending on time and space. In (\ref{RANS2}), the terms $-\rho \overline{u_i'u_j'}$ give the Reynolds stresses calculated by using the Boussinesq hypothesis to close the RANS equations as given by \cite{cebeci2005computational} \cite{fluent202112}  
\begin{equation}
	-\rho \overline{u_i'u_j'} = \mu_t \left( \frac{\partial \overline{u_i}}{\partial x_j} + \frac{\partial \overline{u_j}}{\partial x_i} \right) - \frac{2}{3} \left(\rho k + \mu_t \frac{\partial \overline{u_k}}{\partial x_k} \right)
\end{equation}
where $\mu_t$ is the turbulent viscosity and $k$ is the turbulent kinetic energy. Here, $k$ and its dissipation rate ($\epsilon$) are obtained by the realizable $k-\epsilon$ model which is a widely used and accurate turbulence model as applied in \cite{busco2020sneezing}. Although LES model is more accurate than $k-\epsilon$ model, it is computationally very complex and not appropriate for our aim which is to obtain statistical data by running several simulations. Transport equations for realizable $k-\epsilon$ model are given by \cite{shih1995new}
\begin{equation}
	\frac{\partial (\rho k)}{\partial t} \hspace{-0.05cm} + \hspace{-0.05cm} \frac{\partial (\rho k u_j)}{\partial x_j} \hspace{-0.05cm} = \hspace{-0.05cm} \frac{\partial}{\partial x_j} \left[ \left( \mu \hspace{-0.05cm} + \hspace{-0.05cm} \frac{\mu_t}{\sigma_k}  \right) \frac{\partial k}{\partial x_j}\right] \hspace{-0.05cm} + \hspace{-0.05cm} G_k \hspace{-0.05cm} - \hspace{-0.05cm}\rho \epsilon \hspace{-0.05cm} + \hspace{-0.05cm} S_k \label{k-ep1} \vspace{-0.5cm}
\end{equation}
\begin{multline}
	\frac{\partial (\rho \epsilon)}{\partial t} \hspace{-0.05cm} + \hspace{-0.05cm} \frac{\partial (\rho \epsilon u_j)}{\partial x_j} \hspace{-0.05cm} = \hspace{-0.05cm} \frac{\partial}{\partial x_j} \left[ \left( \mu + \frac{\mu_t}{\sigma_{\epsilon}}  \right) \frac{\partial \epsilon}{\partial x_j}\right] +  \rho C_1S \epsilon \\ - \rho C_2 \frac{\epsilon^2}{k+\sqrt{\nu \epsilon}} + S_{\epsilon}, \label{k-ep2}
\end{multline}
where $C_1 = \max\left[0.43, \frac{\eta}{\eta+5}\right]$, $\eta = S k/\epsilon$, $S$ is the mean rate of strain tensor $S = \sqrt{2 S_{ij}S_{ij}}$ with $S_{ij}$ being the strain tensor, $S_k$ and $S_{\epsilon}$ are user-defined source terms, $C_2 = 1.9$, $C_{1 \epsilon} = 1.44$, $\sigma_k = 1$ and $\sigma_{\epsilon} = 1.2$ are the turbulent Prandtl numbers for $k$ and $\epsilon$, respectively. Here, $G_k$ which is the term related with the turbulent kinetic energy is given as $G_k = \mu_t S^2$ with $\mu_t$ as given by
\begin{equation}
	\mu_t = \rho C_{\mu}\frac{k^2}{\epsilon} \label{mu_t}
\end{equation}
where $C_{\mu}$ is calculated by the formulas given in \cite{shih1995new} (equations (19)-(21) in \cite{shih1995new}).

By solving the equations (\ref{RANS1})-(\ref{mu_t}) iteratively, the average airflow velocities can be obtained whereas the fluctuating airflow velocities are calculated as follows.

It should be noted that a constant airflow velocity ($v_{air}$) is added to $\overrightarrow{u}$ in the $+z$ direction towards the RX in addition to the turbulent airflow velocity as also shown in Fig. \ref{Geo}. The stochastic effect of turbulence is incorporated to the system model by adding the fluctuation  values, i.e., $u_i'$, via the discrete random walk (DRW) model so that the turbulent dispersion of particles are modeled. According to DRW model, $u_i'$ is determined as \cite{mofakham2020improved} \cite{fluent202112}
\begin{equation}
	u_i' = \beta \sqrt{\frac{2 k}{3}}
\end{equation}
where $\beta$ is a standard normal random variable, i.e., $\beta = \mathcal{N}(0,1)$. Each particle is tracked along the eddy interaction time as given by $t_{int} = \min(\tau_e, t_R)$ where $\tau_e$ is the eddy lifetime in a turbulent flow and calculated by $\tau_e = -T_L \ln(r)$ where $r$ is a standard uniform random variable and $T_L \approx 0.15 k/ \epsilon$ is the Lagrangian integral time. In addition, $t_R$ is the particle eddy transit (or crossing) time as given by
\begin{equation}
	t_R = -\tau_r \ln\left[1 - \left(\frac{L_e}{\tau_r|\overrightarrow{u}-\overrightarrow{u_p}|} \right) \right],
\end{equation}
where $L_e$ is the eddy length scale, $\overrightarrow{u}$ and $\overrightarrow{u_p}$ are the air and particle velocities, respectively. Here, $\tau_r = (24 \rho_p d_p^2)/(18 \mu C_D Re)$ is the particle relaxation time \cite{gordon1968error}.  A new value is assigned to $u_i'$ via updating $\beta$, when $t_{int}$ is reached during the tracking of a particle. At each time step, $\tau_e$ is updated according to the changing $k$ and $\epsilon$ at each point in the flow domain.

\subsubsection{Discrete Phase} \label{DP}
According to the Newton's second law of motion, the acting forces on a spherical particle is given by
\begin{equation}
	m_p \frac{d \overrightarrow{u_p}}{dt} = m_p \frac{\overrightarrow{u}-\overrightarrow{u_p}}{\tau_r} + m_p \frac{\overrightarrow{g}(\rho_p-\rho)}{\rho_p},
\end{equation}
where $m_p$ shows the particle mass, $\rho_p$ is the particle density, and $\overrightarrow{g}$ is the gravitational acceleration. Here, the second term on the right hand side shows the net force downwards (difference between gravitational and buoyant force) as also derived in \cite{gulec2021molecular}, and the first term on the right hand side gives the drag force, $Re$ is the Reynolds number, $d_p$ is the particle diameter. $C_D$ is the drag coefficient following the spherical drag law as given by \cite{morsi1972investigation}
\begin{equation}
	C_D = K_1 + \frac{K_2}{Re} + \frac{K_3}{Re^2}
\end{equation}
where $K_1$, $K_2$ and $K_3$ are experimentally validated constants which change according to  $Re$ as given in \cite{morsi1972investigation}.

\begin{figure*}[t]
	\vspace{0.13in}
	\centering
	\includegraphics[width=0.8\textwidth]{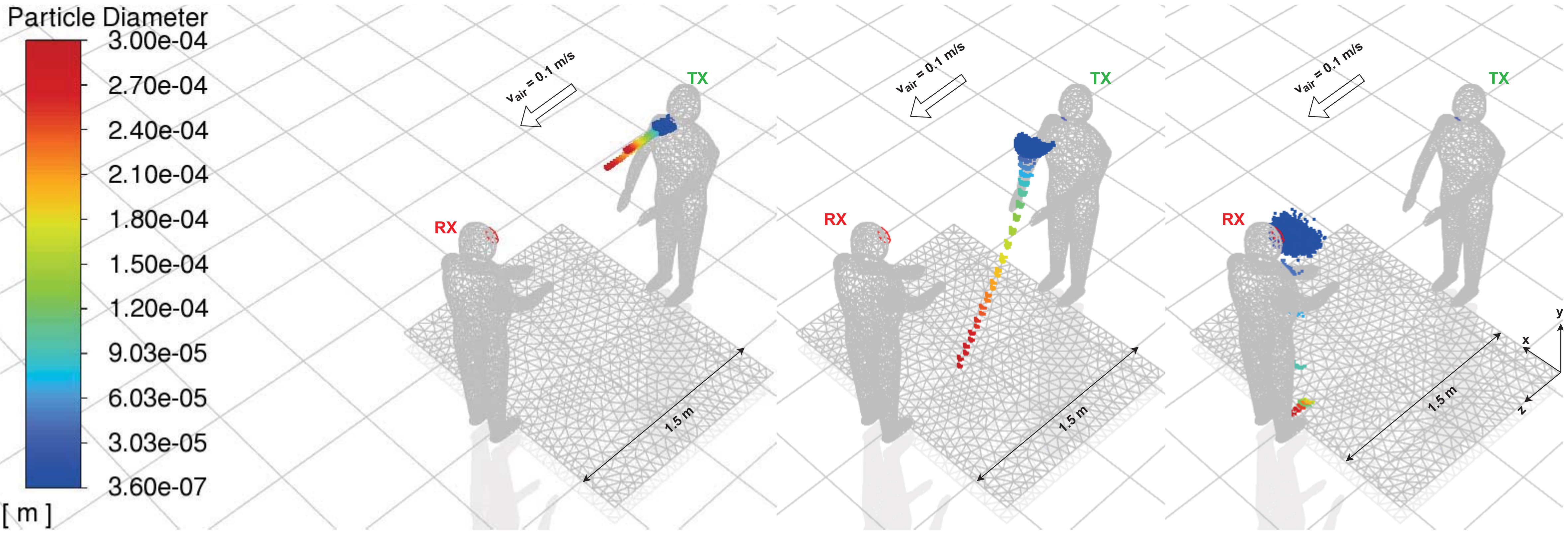} \\
	\scriptsize \hspace{1.45 in}   (a) \hspace{1.3 in}  (b) \hspace{1.25 in} (c) \\
	\caption{Visualization of a human cough and its reception with 48800 particles for $v_{air} = 0.1$ m/s at (a) $t = 0.12$ s (b) $t = 1.5$ s (c) $t = 12$ s.}
	\label{Flow}
	\vspace{-.5em}
\end{figure*}

\begin{figure*}[t]
	\vspace{0.13in}
	\centering
	\includegraphics[width=0.8\textwidth]{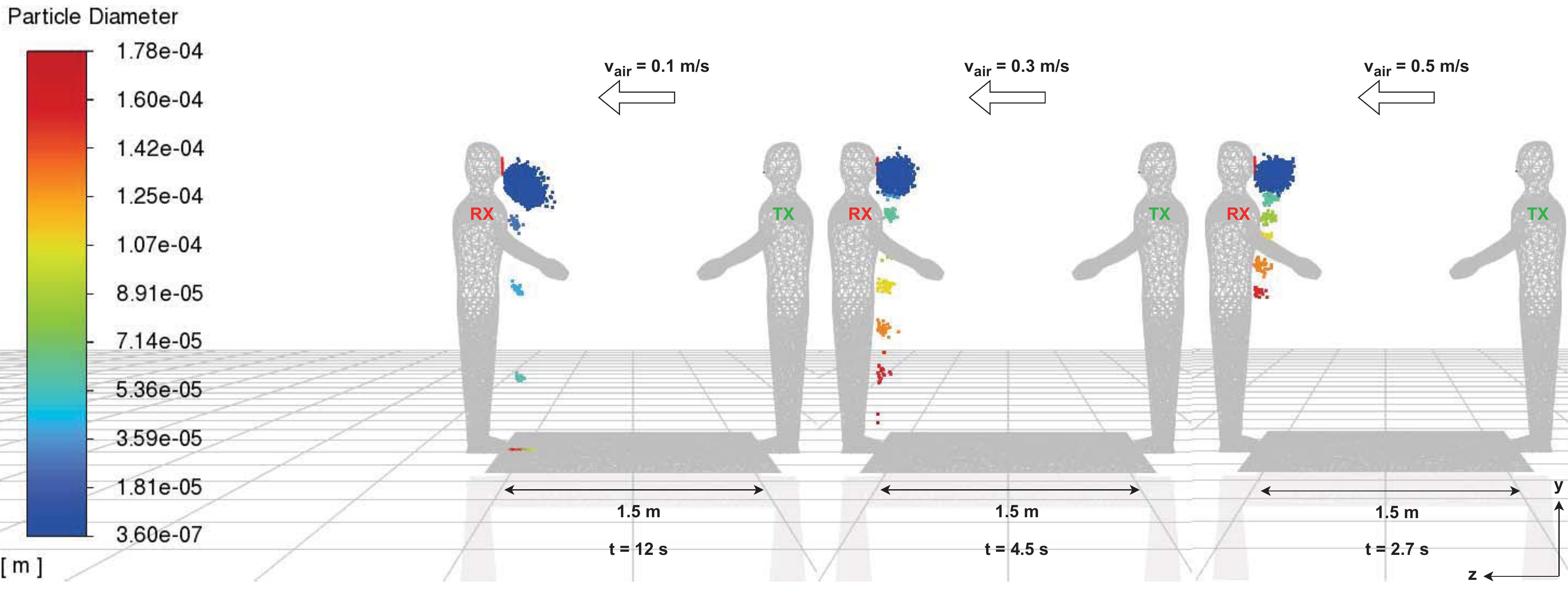} \\
	\scriptsize \hspace{1.6 in}   (a) \hspace{1.2 in}  (b) \hspace{1.2 in} (c) \\
	\caption{The effect of different air velocities and turbulence on the pathogen-laden cough particles during the reception for  (a) $v_{air} = 0.1$ m/s at $t = 12$ s (b) $v_{air} = 0.3$ m/s at $t = 4.5$ s (c) $v_{air} = 0.5$ m/s at $t = 2.7$ s.}
	\label{Reception}
\end{figure*}

\subsection{Receiver Model}
As for the reception, particles reaching at the RX cross-section as shown in Fig. \ref{Geo} are counted until every particle leaves the flow domain. This RX cross-section is in the center of the human face including eyes, nose and the mouth and is a circle having a diameter $r_R = (\sqrt{\beta_{bb}^2 + \beta_{ss}^2})/2$ where $\beta_{bb}$ is the biocular breadth and $\beta_{ss}$ is the Sellion-Stomion length as also given in \cite{gulec2021molecular}. Next, the usage of the models by numerically solving the equations in this section in a CFD simulator is elaborated.

\section{Computational Fluid Dynamics Simulator} \label{CFD_Setup}
In this section, the details about the setup of CFD simulations which are executed by using Ansys Fluent 2021 R.2 simulator are given. As shown in Fig. \ref{Geo}, two identical manikins are used as the TX and RX which are $176$ cm high. The flow domain which is shown as a green cuboid has the 3-D dimensions $2\times2\times1.5$ m. The airflow and particle motions are simulated within this flow domain which includes the TX emission and RX reception surfaces. The CFD simulator calculates variables such as pressure and velocity with the finite volume method which requires the flow domain to be divided into small volumes. To this end, meshing is performed by generating $74520$, $25772$ and $26092$ tetrahedral cells for the flow domain, TX and RX volumes, respectively. 


The mouth of the TX is modeled as an ellipse ($4 \times 1$ cm, area $=314$mm$^2$) \cite{pendar2020numerical} and aligned in the same axis with the RX. The center of the RX cross-section and TX are at a height of $162.6$ cm and $159$ cm high from the ground, respectively. Particles and air are emitted with an initial velocity ($u_0$) from this mouth surface along the emission time ($T_e$) according to the size distributions derived in Section \ref{Particle_Distr}. 

In the flow domain given in in Fig. \ref{Geo}, boundary conditions are arranged as follows. The rectangular surface ($z = 1.5$ m) where the RX circular cross-section is also deployed is configured as an absorbing boundary in order to count the received number of particles. All the other surfaces of the cuboid and the human body surfaces are configured as reflecting surfaces. When there is no constant airflow ($v_{air}$) except the initial impulsive cough velocity, the boundary conditions can be much more effective on the propagation of particles. In addition, the simulation of this scenario without considering $v_{air}$ takes more time, which is not feasible for our purpose of observing the statistical distribution of received particles. Furthermore, real life scenarios which include more reflecting or absorbing boundary conditions such as offices or supermarkets can be more complicated. However, these scenarios are beyond the scope of this paper.

By using these boundary conditions and the governing equations (\ref{RANS1})-(\ref{k-ep2}), the CFD simulator discretize these equations by converting them from their integral form to algebraic equations relating $p$ and $\overline{u_i}$ values at each cell center \cite{fluent202112}. Then, these equations are linearized via Taylor series expansions and solved iteratively by guessing the pressure and velocity values at each cell center after each iteration. At each iteration, the mass conservation is calculated and iterations continue until the error of mass imbalance converges. In Ansys Fluent, the coupled  pressure-based solver is used for the calculations of transient simulations. After $\overline{u_i}$ is determined for each cell at each time step ($\Delta t$), these values are used by also using the models explained in Section \ref{DP} to calculate the turbulent dispersion of particles (including their interactions with continuous phase), i.e., their instantaneous velocities and positions. 

\begin{table}[b]
	\vspace{-1em}
	\centering
	\caption{Simulation parameters}
	\scalebox{0.85}{
		\begin{tabular}{p{45pt}|p{81pt}|p{43pt}|p{74pt}}
			\hline
			\textbf{Parameter}	& \textbf{Value} & \textbf{Parameter}	& \textbf{Value}\\
			\hline 
			$\Delta t$  &  $ 0.06$ s & $t_s$ & $\{18, 9, 6\}$ s \\
			$T_e$ &$ 0.12 $ s \cite{scharfman2016visualization} & $u_0$ & $11.2$ m/s (cough) \cite{zhu2006study}\\
			$ Q_d $ & $47.83\times10^-6$ kg/s \cite{xie2009exhaled} & $g$ & $9.81$ m/s$^2$  \\
			$Q_a$ &  $9.9259$ kg/s \cite{lindsley2012quantity} & $\mu$ & $17.894\hspace{-0.07cm} \times \hspace{-0.07cm} 10^{-6}$ kg/(m s) \\
			$\rho$  & $1.225$ kg/m$^3$  & $\beta_{bb}$ (female) & $8.853$ cm \cite{young1993head} \\
			$ \rho_p $ & $998.2$ kg/m$^3$ & $\beta_{bb}$ (male) & $9.131$ cm  \cite{young1993head} \\
			$N_d$ & $800$ & $\beta_{ss}$ (female) &$6.901$ cm \cite{young1993head}\\
			$N_a$  & $48000$ & $\beta_{ss}$ (male) & $7.57$ cm \cite{young1993head}\\	
			\hline           
	\end{tabular}}
	\label{Sim_parameters}
\end{table}

\begin{figure*}[tb]
	\centering
	\includegraphics[width=0.325\textwidth]{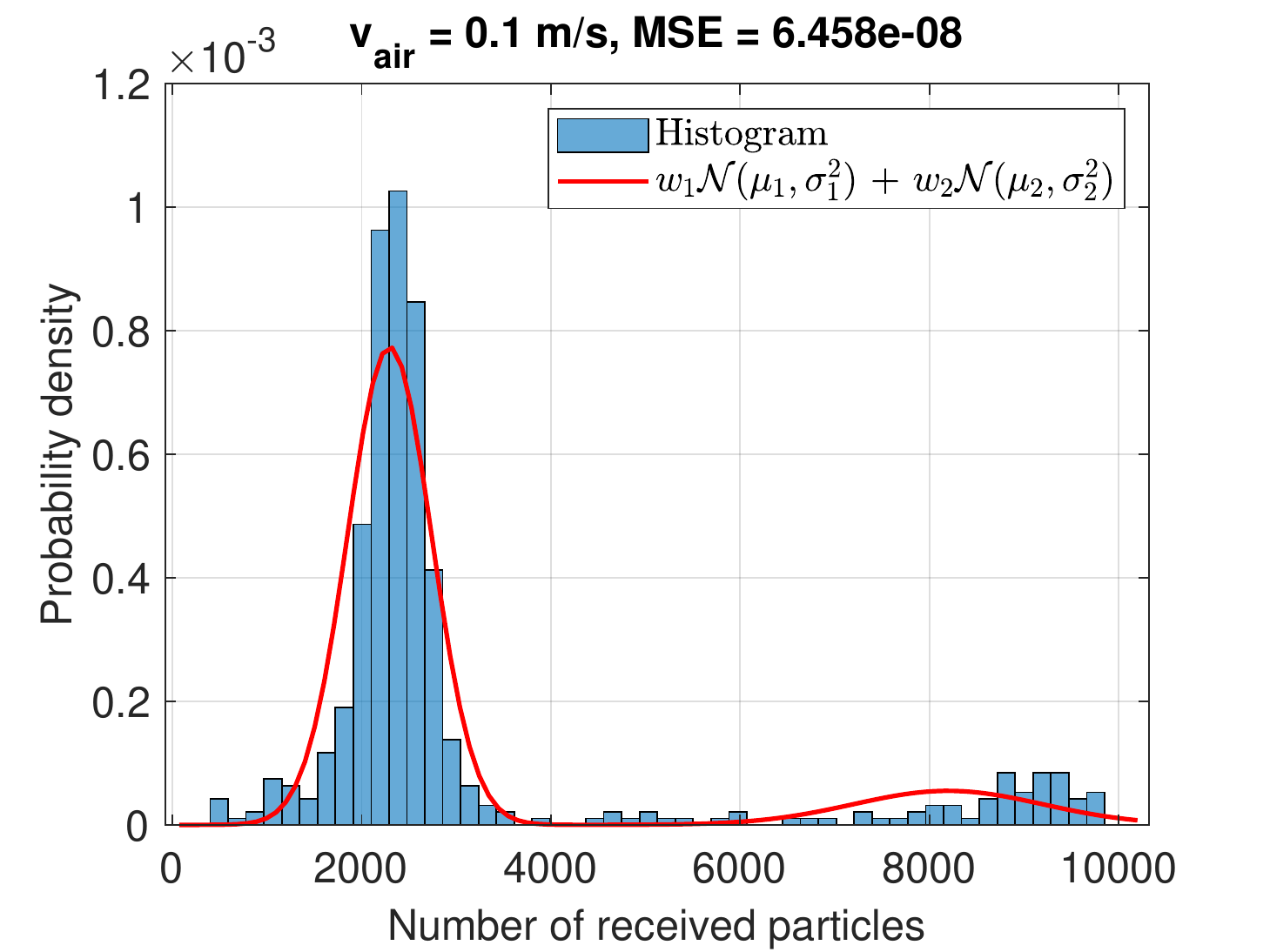}    
	\includegraphics[width=0.325\textwidth]{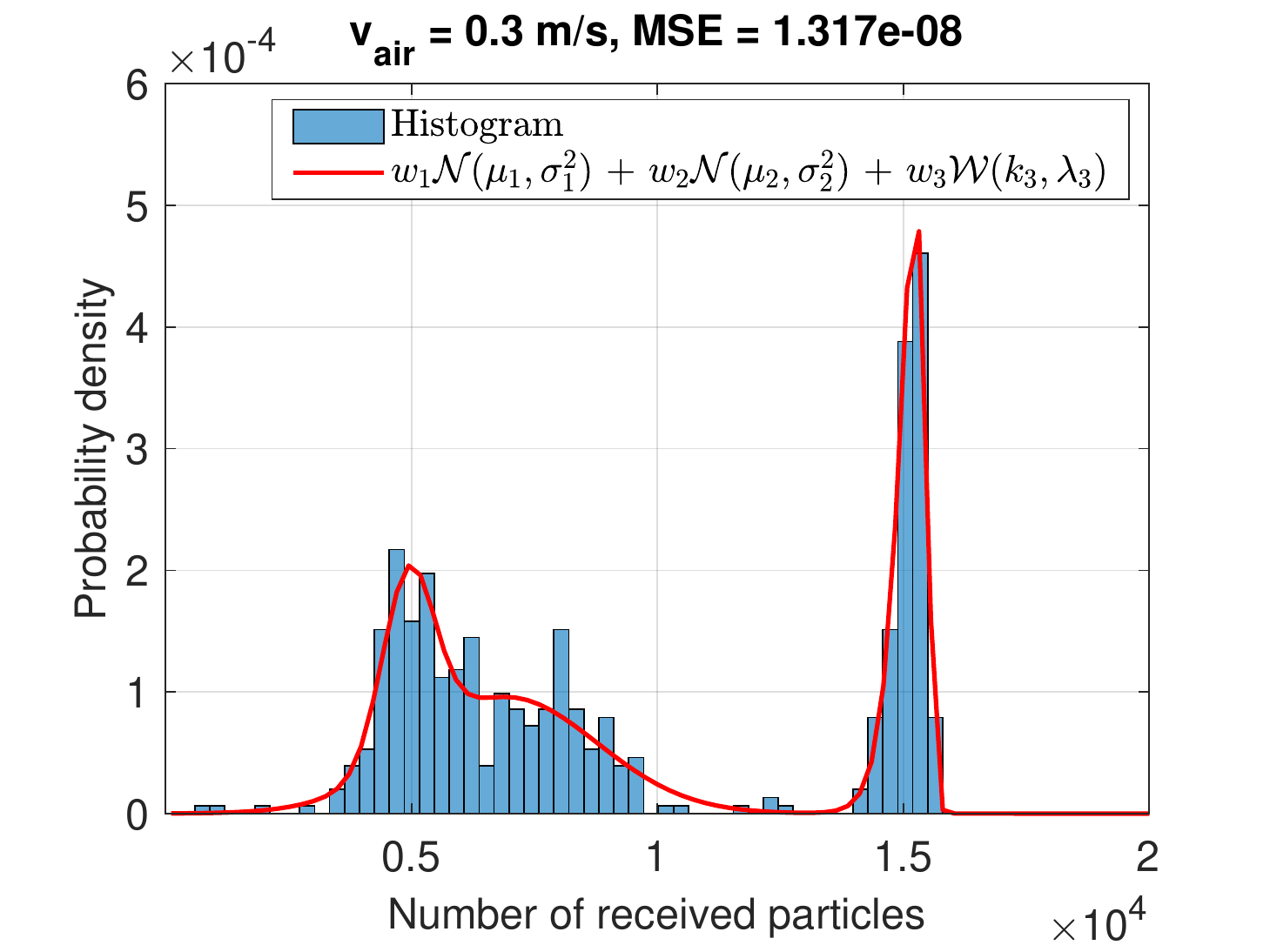}  
	\includegraphics[width=0.325\textwidth]{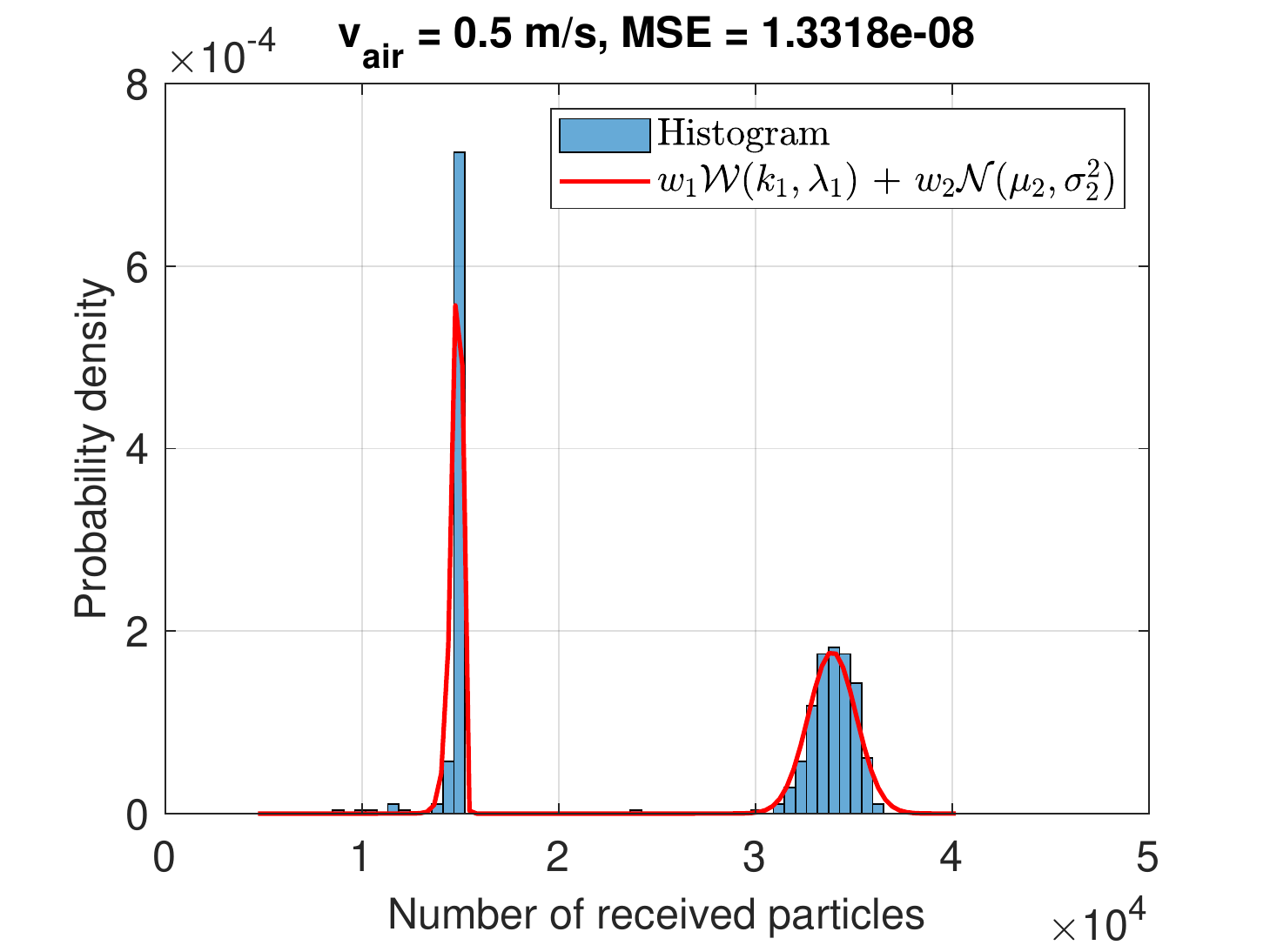} \\
	\scriptsize \hspace{0.1 in}  (a) \hspace{2.2 in}  (b) \hspace{2.2 in} (c) \\
	\caption{Probability density function of $ N_R $ for (a) $v_{air} = 0.1$ m/s (b) $v_{air} = 0.3$ m/s (c) $v_{air} = 0.5$ m/s.}
	\label{pdf123}
\end{figure*}

\begin{table*}[tb]
	\centering
	\caption{Estimated statistical parameters of $ f_{N_{R_i}} $ }
	\scalebox{1}{
		\begin{tabular}{p{15pt}|p{10pt}|p{40pt}|p{40pt}|p{10pt}|p{40pt}|p{40pt}|p{10pt}|p{40pt}|p{40pt}|p{40pt}} \hline
			Pdf	& $ w_1 $ & \multicolumn{2}{|l|}{Distribution 1}                                   & $ w_2 $ & \multicolumn{2}{|l|}{Distribution 2}                                  & $ w_3 $ & \multicolumn{2}{|l|}{Distribution 3}                & MSE      \\ \hline
			$ f_{N_{R_1}} $ & $ 0.86 $ & $\mu_1$=$2294.9$ & $\sigma_1$=$444.2$ & $ 0.14 $ & $\mu_2$=$8195.7 $  & $ \sigma_2$=$1000 $   & -    & -           & -                                   & $ 6.458\mathrm{e}{-8} $ \\ \hline
			$ f_{N_{R_2}} $ & $ 0.20 $ & $\mu_1$=$4938.1$ & $\sigma_1$=$527.5$ & $ 0.44 $ &$\mu_2$=$6929.5 $  & $ \sigma_2$=$1846.1 $ & $ 0.36 $ & $ k_3$=$58.4 $ & $ \lambda_3$=$15242.7 $ & $ 1.317\mathrm{e}{-8} $ \\ \hline
			$ f_{N_{R_3}} $ & $ 0.46 $ & $k_1$=$60.9$                    &$  \lambda_1$=$14915.6 $ & $ 0.54 $ & $ \mu_2$=$33888.1 $ & $ \sigma_2$=$1218.8 $ & -    & -           & -                                   & $ 1.332\mathrm{e}{-8} $ \\ \hline
	\end{tabular}}
	\label{Est_parameters}
\end{table*}

\section{Simulation Results} \label{SR}
In this section, visual simulation results which are obtained via the CFD simulator as detailed in Section \ref{CFD_Setup} are presented. The CFD simulations and the observations for the reception, i.e., counting particles at the RX, are repeated $500$ times for each of three different air velocities, i.e., $v_{air} = \{0.1, 0.3, 0.5\}$ m/s using the parameter values in Table \ref{Sim_parameters} along $t_s$. These chosen  air velocities are based on still air and two different ventilation scenarios similar to the works in \cite{vuorinen2020modelling, dbouk2020coughing}. In indoor environment, there is nearly always a slight airflow, even if it seems like still air \cite{gulec2020localization}. The scenario when $v_{air} = 0.1$ m/s corresponds to the indoor still air environment. The other velocities ($0.3$, $0.5$ m/s) represent two different ventilation scenarios induce by open doors, windows or ventilation systems. The initial cough flow rates for droplets ($ Q_d $) and aerosols ($ Q_a $) are calculated by dividing the emitted droplet and aerosol  masses to $T_e$. The number of emitted droplets ($N_d$) and aerosols ($N_a$) are given in Table \ref{Sim_parameters}. Here, the numbers of each particle size is distributed according to the Weibull distribution values estimated in Section \ref{Particle_Distr}. In addition, $\beta_{bb}$ and $\beta_{ss}$ values are taken as the average value of female and male values to calculate $r_R$ of the RX as also applied in \cite{gulec2021molecular}.

In Fig. \ref{Flow}, the trajectory and dispersion of the cough particles can be observed for $v_{air} = 0.1$ m/s between the emission and reception. At the initial state, larger particles tend to move faster due to the initial velocity as shown in Fig. \ref{Flow} (a). However, as these larger particles (or large droplets) continue their movement, they are affected by the air drag and lose their momentum more quickly. In addition, large droplets fall down to the ground due to the gravity as it is clearly observed in Fig. \ref{Flow} (b). In contrast to large droplets, aerosols are affected less by the gravity and air drag due to their small sizes. Therefore, they are entrained by the ambient air flow towards the RX and large droplets cannot reach at the RX for a distance of $1.5$ m. Furthermore, the effect of ambient air velocity on the dispersion of particles is shown in Fig. \ref{Reception}. These results show that as $v_{air}$ increases, particles propagate for a longer distance before they fall down to the ground. Besides, smaller $v_{air}$ causes more dispersion on the particles as shown in Fig. \ref{Reception} (a)-(c), while it increases the reception time of droplets. The reason of this dispersion is the randomness due to the effect of turbulence at the initial state of the particle propagation. In the next section, data obtained via CFD simulations are used for the statistical characterization of the probability of infection.

\section{Probability of Infection}\label{PoI}
In this section, the collected data for the received number of particles via the CFD simulations are analyzed and employed for statistical characterization of the reception in airborne pathogen transmission. The probability of infection is derived for different scenarios by using these statistics and an analysis is given based on the reception statistics. 

\begin{figure*}[bt]
	\centering
	\includegraphics[width=0.325\textwidth]{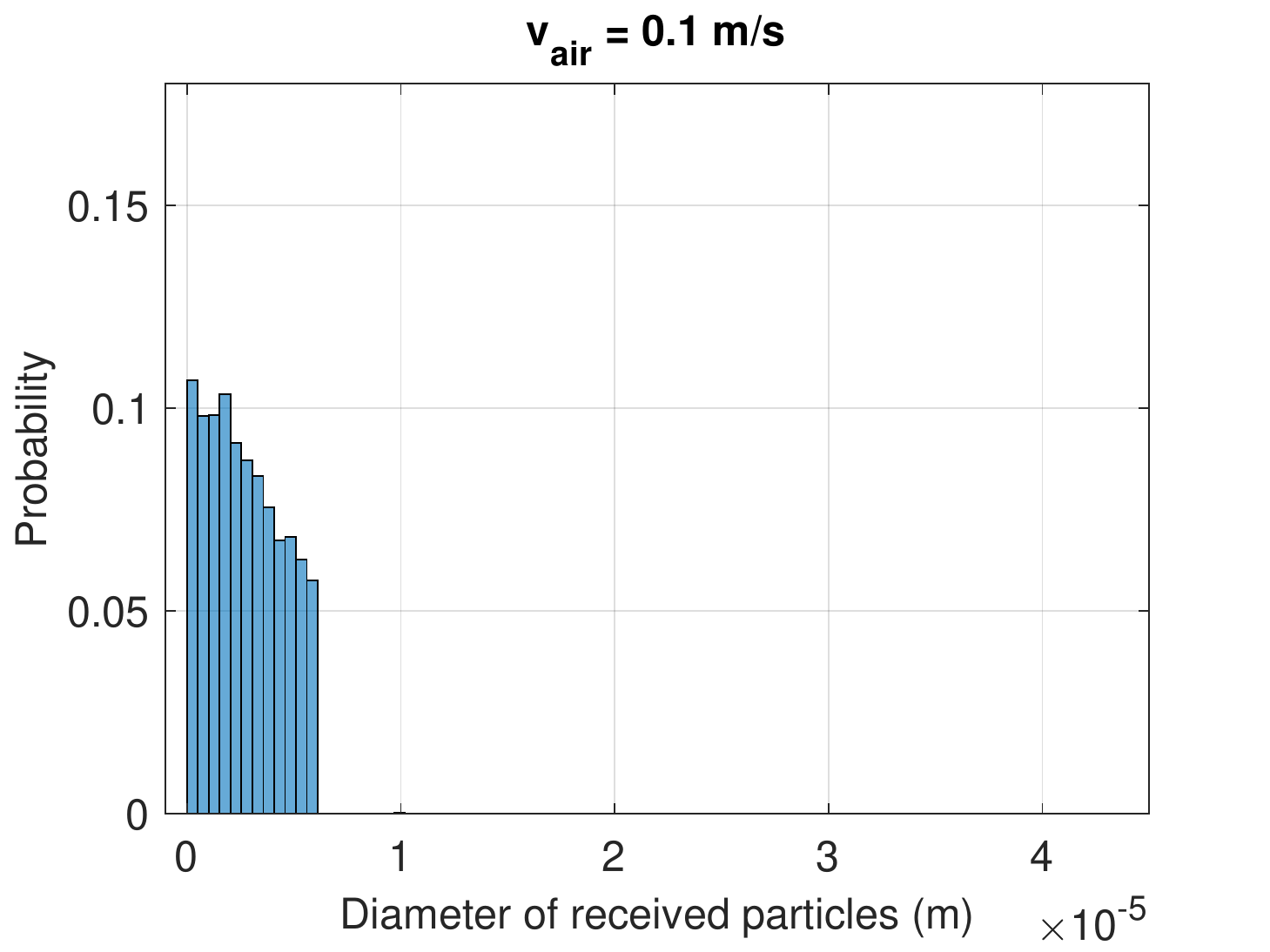}    
	\includegraphics[width=0.325\textwidth]{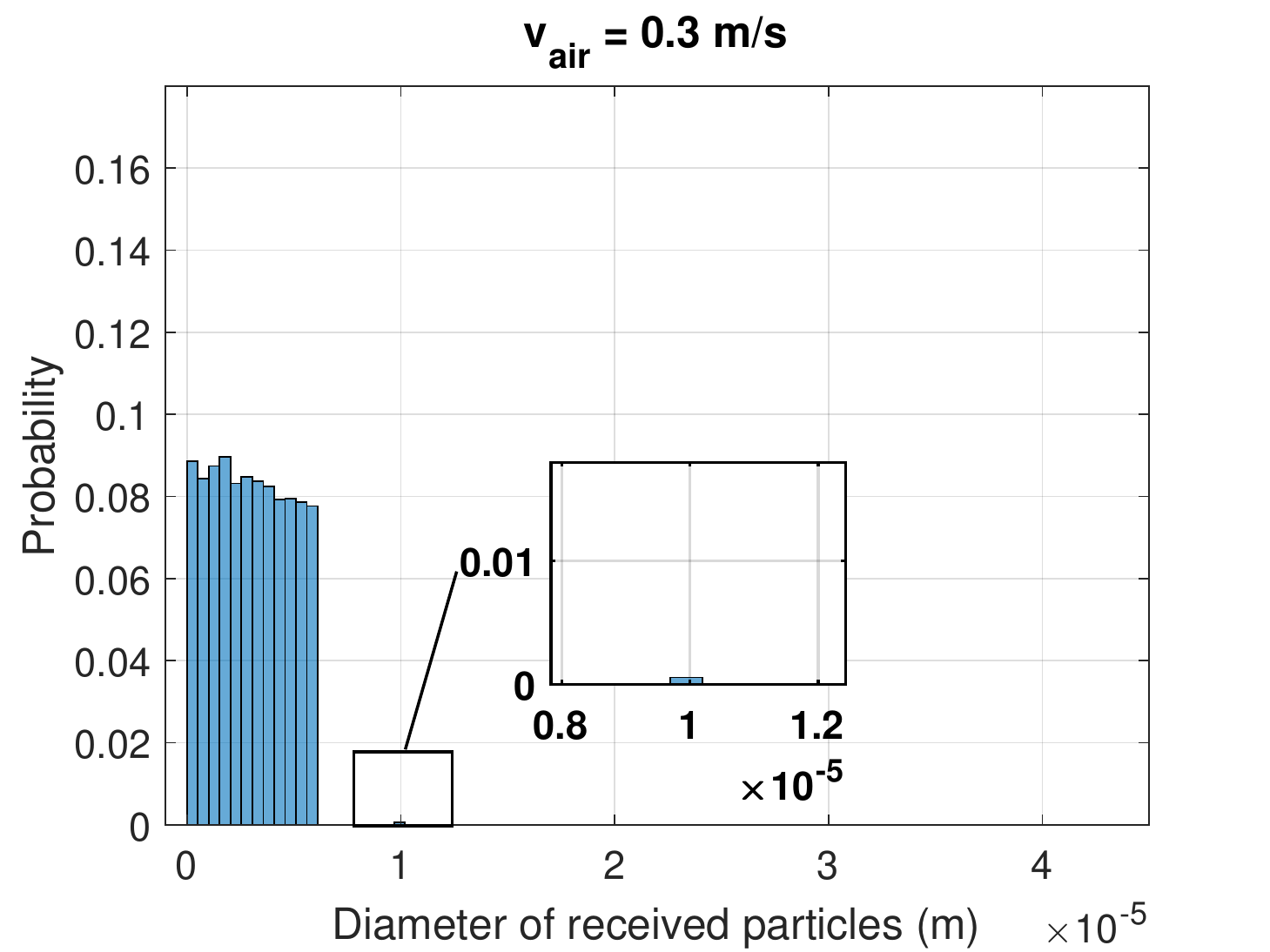}  
	\includegraphics[width=0.325\textwidth]{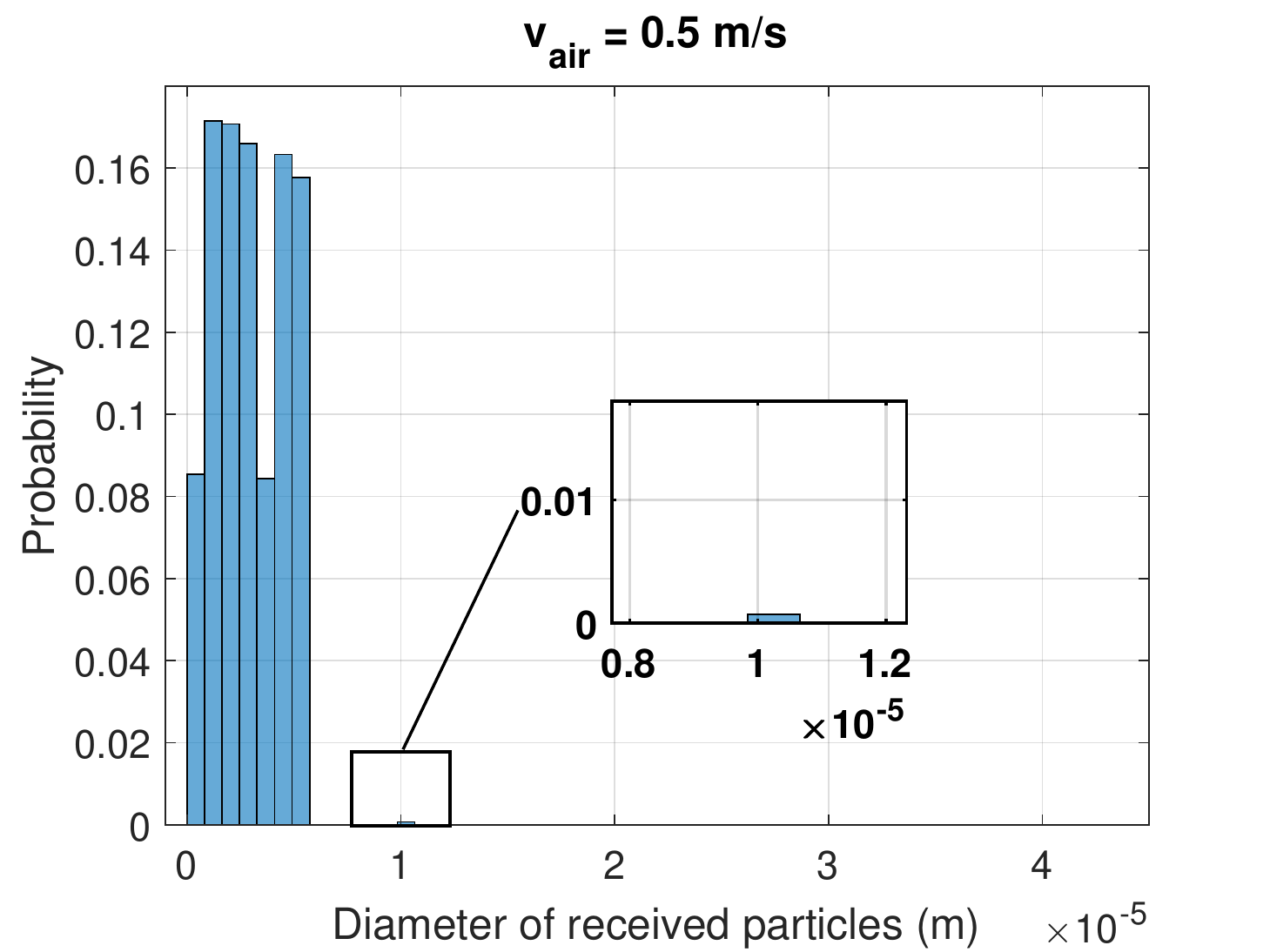} \\
	\scriptsize \hspace{0.1 in}  (a) \hspace{2.2 in}  (b) \hspace{2.2 in} (c) \\
	\caption{Probability of diameters of received particles for (a) $v_{air} = 0.1$ m/s (b) $v_{air} = 0.3$ m/s (c) $v_{air} = 0.5$ m/s.}
	\label{Dia}
\end{figure*}

\subsection{Characterization}
In Fig. \ref{pdf123}, the histograms and the fitted probability density functions (pdfs) of the received number of particles for three different ambient air velocities are given. As observed from this figure, the distributions are multi-modal and can be modeled by using the weighted sums of different distributions. To this end, these weighted distributions are determined visually according to the shapes of histograms, which are given as 
\begin{align}
\hspace{-0.2cm}	f_{N_{R_1}}(N_R) \hspace{-0.05cm} &= \hspace{-0.05cm} w_1 \mathcal{N}(\mu_1,\sigma_1^2) \hspace{-0.1cm} + \hspace{-0.1cm} w_2 \mathcal{N}(\mu_2,\sigma_2^2) \label{pdf_eq1}\\
\hspace{-0.2cm}	f_{N_{R_2}}(N_R) \hspace{-0.05cm} &= \hspace{-0.05cm} w_1 \mathcal{N}(\mu_1,\sigma_1^2) \hspace{-0.1cm} + \hspace{-0.1cm} w_2 \mathcal{N}(\mu_2,\sigma_2^2) \hspace{-0.1cm} + \hspace{-0.1cm} w_3 \mathcal{W}(k_3,\lambda_3)  \label{pdf_eq2} \\
\hspace{-0.2cm}	f_{N_{R_3}}(N_R) \hspace{-0.05cm} &= \hspace{-0.05cm} w_1 \mathcal{W}(k_1, \lambda_1) \hspace{-0.1cm} + \hspace{-0.1cm} w_2 \mathcal{N}(\mu_2, \sigma_2^2),
	\label{pdf_eq3}
\end{align}
where $w_1$, $w_2$ and $w_3$ show the weights, $ \mathcal{N}(\mu_R, \sigma_R^2) $ is the normal distribution with mean $\mu_R$ and variance $\sigma_R^2$, $ \mathcal{W}(k_R, \lambda_R) $ is the Weibull distribution with the scale parameter $\lambda_R$ and shape parameter $k_R$, and $ f_{N_{R_1}}(N_R) $, $ f_{N_{R_2}}(N_R) $, $ f_{N_{R_3}}(N_R) $ are the pdfs for $v_{air} = 0.1$ m/s, $v_{air} = 0.3$ m/s, and $v_{air} = 0.5$ m/s, respectively. The parameters in (\ref{pdf_eq1})-(\ref{pdf_eq3}), which are estimated via the maximum likelihood estimation method detailed in Section \ref{Particle_Distr}, are given in Table \ref{Est_parameters} with their mean square error (MSE) values.

\begin{figure}[bt]
	\centering
	\includegraphics[width=0.9\columnwidth]{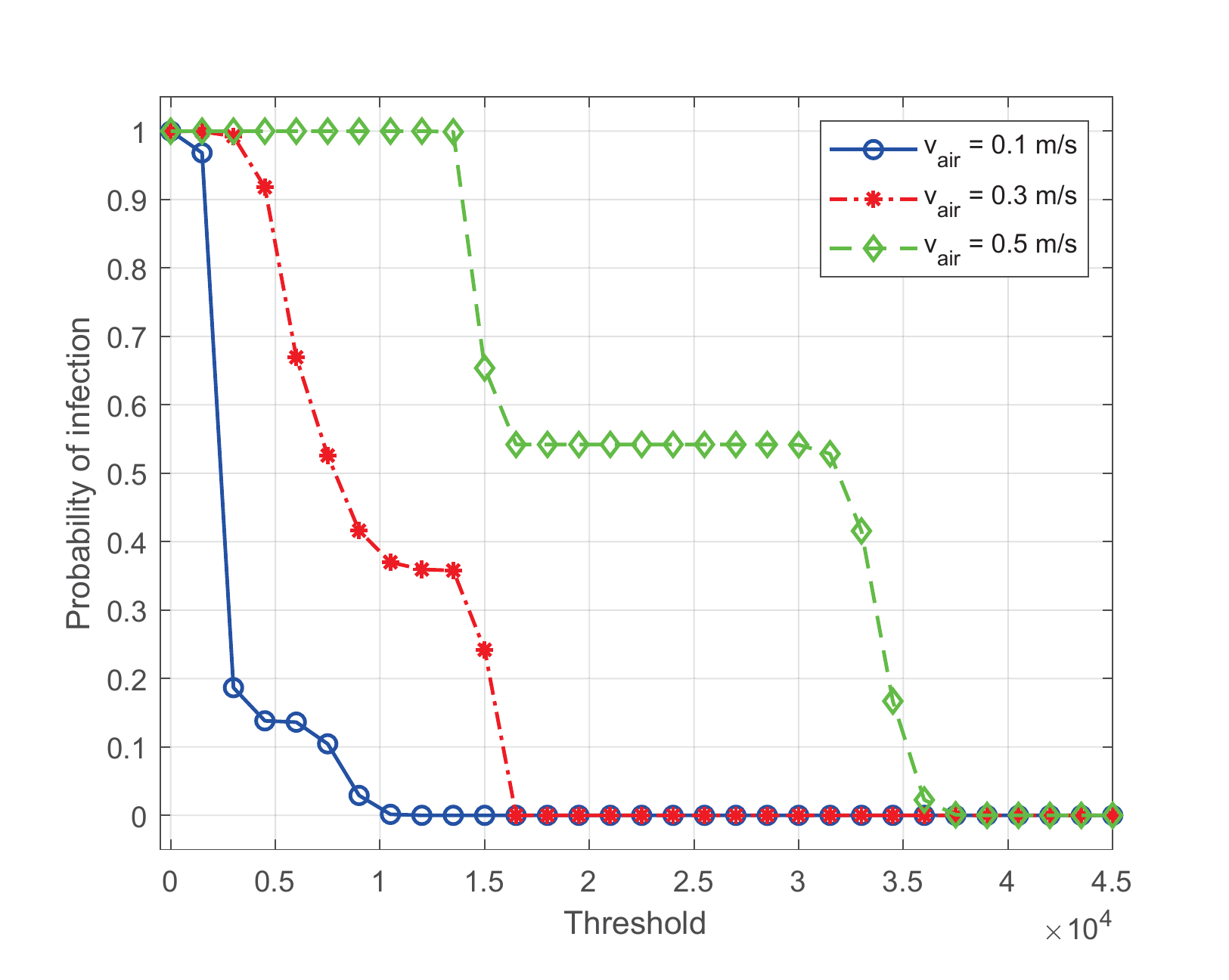}
	\caption{Probability of infection according to $\gamma$ for different $v_{air}$ values.}
	\label{P_inf}
\end{figure}

In the next step, these derived pdfs in (\ref{pdf_eq1})-(\ref{pdf_eq3}) with their estimated parameters given in Table \ref{Est_parameters} can be employed to obtain the probability of infection ($P_{inf}$) as given by \cite{gulec2021molecular}
\begin{equation}
	P_{inf_i} = P(N_R > \gamma) = \int_{\gamma}^{\infty} f_{N_{R_i}}(x) dx \\
	\label{P_infd}
\end{equation}
where $i$ shows the index according to $v_{air}$ as also applied in (\ref{pdf_eq1})-(\ref{pdf_eq3}) and $\gamma$ is the detection threshold corresponding to the immune system's strength of the RX. Thus, the derived expressions for the probability of infections for different air velocities are given by
\begin{align}
\hspace{-0.23cm}	P_{inf_1} \hspace{-0.05cm}  &= \hspace{-0.05cm}  w_1 Q\left(\hspace{-0.05cm}  \frac{\gamma - \mu_1}{\sigma_1} \hspace{-0.05cm} \right) \hspace{-0.05cm}  + \hspace{-0.05cm} w_2 Q\left( \hspace{-0.05cm} \frac{\gamma - \mu_2}{\sigma_2} \hspace{-0.05cm}  \right) \label{P_inf_eq1}\\
\hspace{-0.23cm}	P_{inf_2} \hspace{-0.05cm}  &= \hspace{-0.05cm} w_1 Q\left(\hspace{-0.05cm}  \frac{\gamma - \mu_1}{\sigma_1} \hspace{-0.05cm} \right) \hspace{-0.05cm} + \hspace{-0.05cm} w_2 Q\left(\hspace{-0.05cm}  \frac{\gamma - \mu_2}{\sigma_2} \hspace{-0.05cm}  \right) \hspace{-0.05cm} + \hspace{-0.05cm} w_3 e^{-\left( \frac{\gamma}{\lambda_3} \right)^{k_3}} \label{P_inf_eq2}\\
\hspace{-0.23cm}	P_{inf_3} \hspace{-0.05cm}  &= \hspace{-0.05cm} w_1 e^{- \frac{\gamma}{\lambda_1}^{k_1}} \hspace{-0.05cm} + \hspace{-0.05cm} w_2 Q\left(\hspace{-0.05cm}  \frac{\gamma - \mu_2}{\sigma_2} \hspace{-0.05cm}  \right),
	\label{P_inf_eq3}
\end{align}
where $Q(x) = \frac{1}{\sqrt{2 \pi}} \int_{x}^{\infty} \textrm{e}^{-\frac{u^2}{2}} du$ shows the Q-function.

\begin{figure*}[t]
	\centering
	\includegraphics[width=0.325\textwidth]{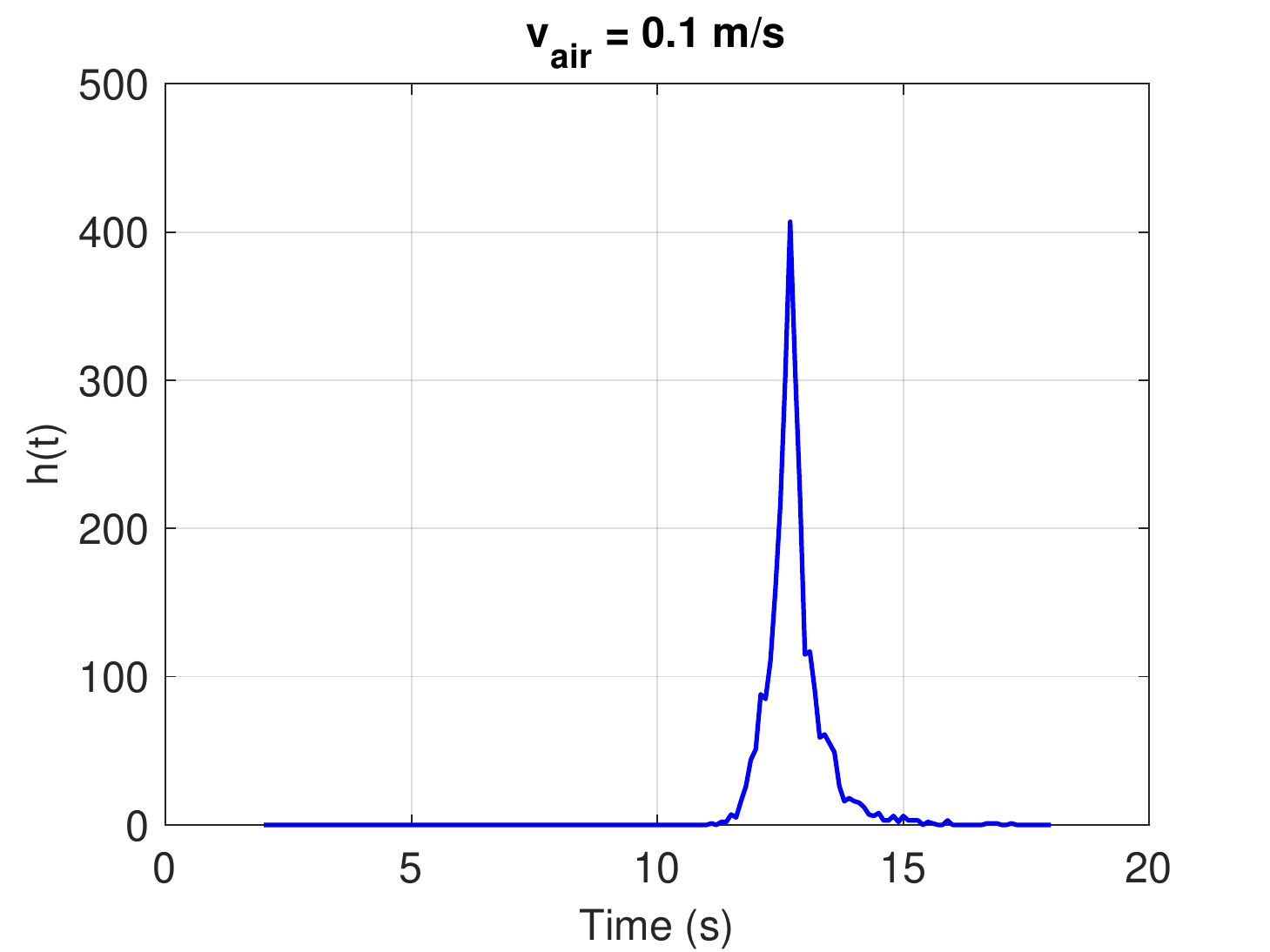}    
	\includegraphics[width=0.325\textwidth]{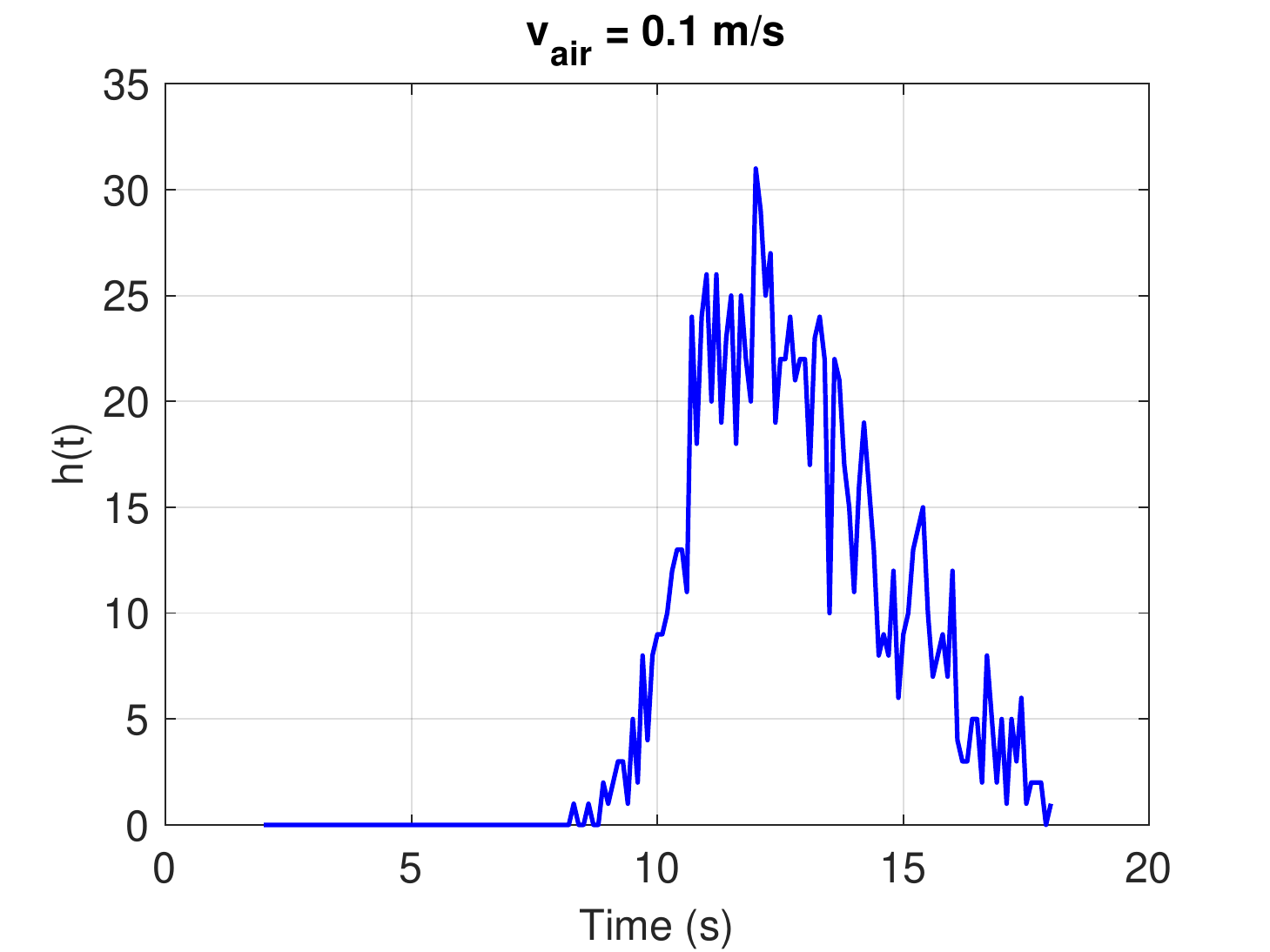}  
	\includegraphics[width=0.325\textwidth]{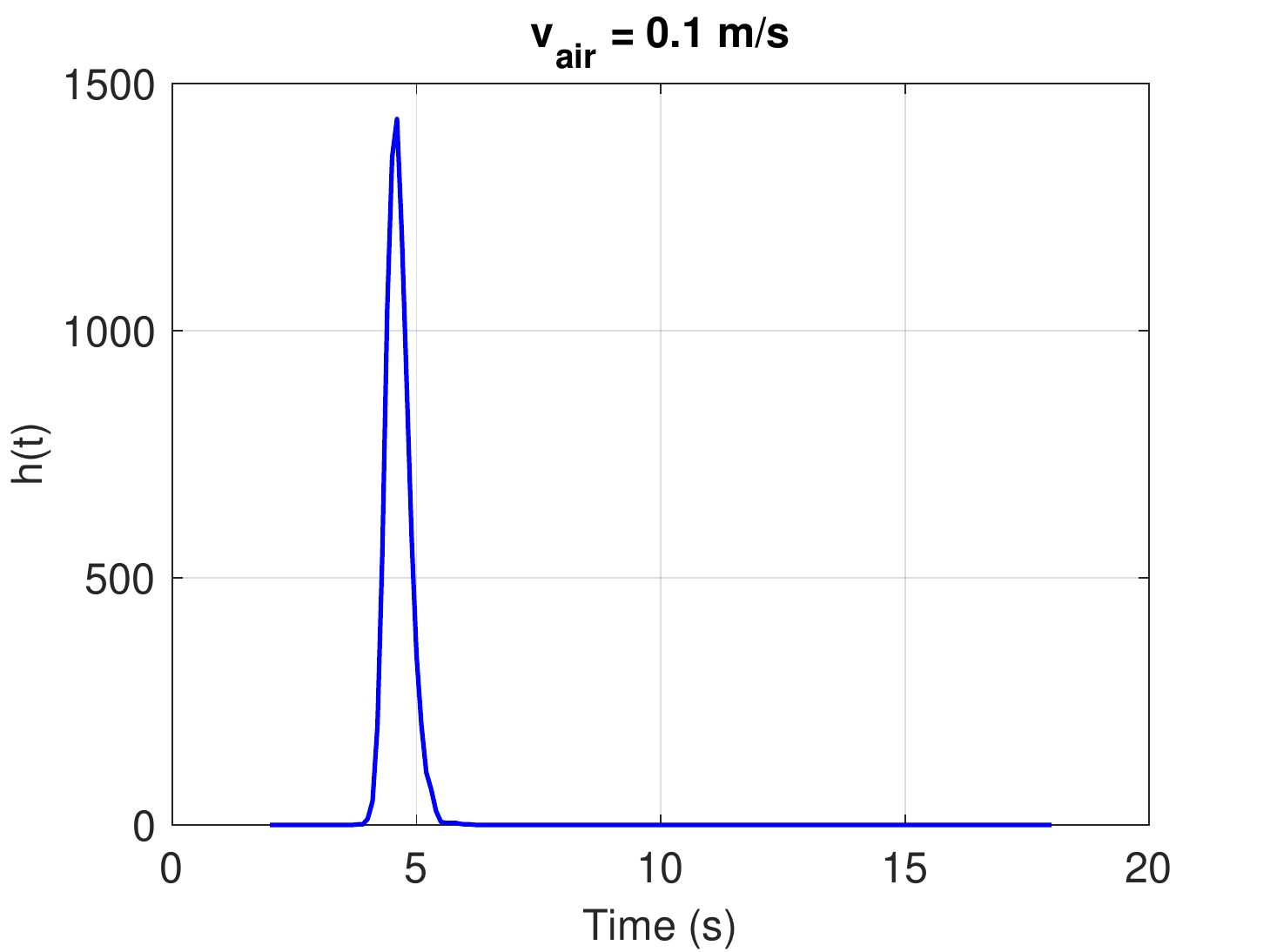} \\
	\scriptsize \hspace{0.1 in}  (a) \hspace{2.2 in}  (b) \hspace{2.2 in} (c) \\
	\includegraphics[width=0.325\textwidth]{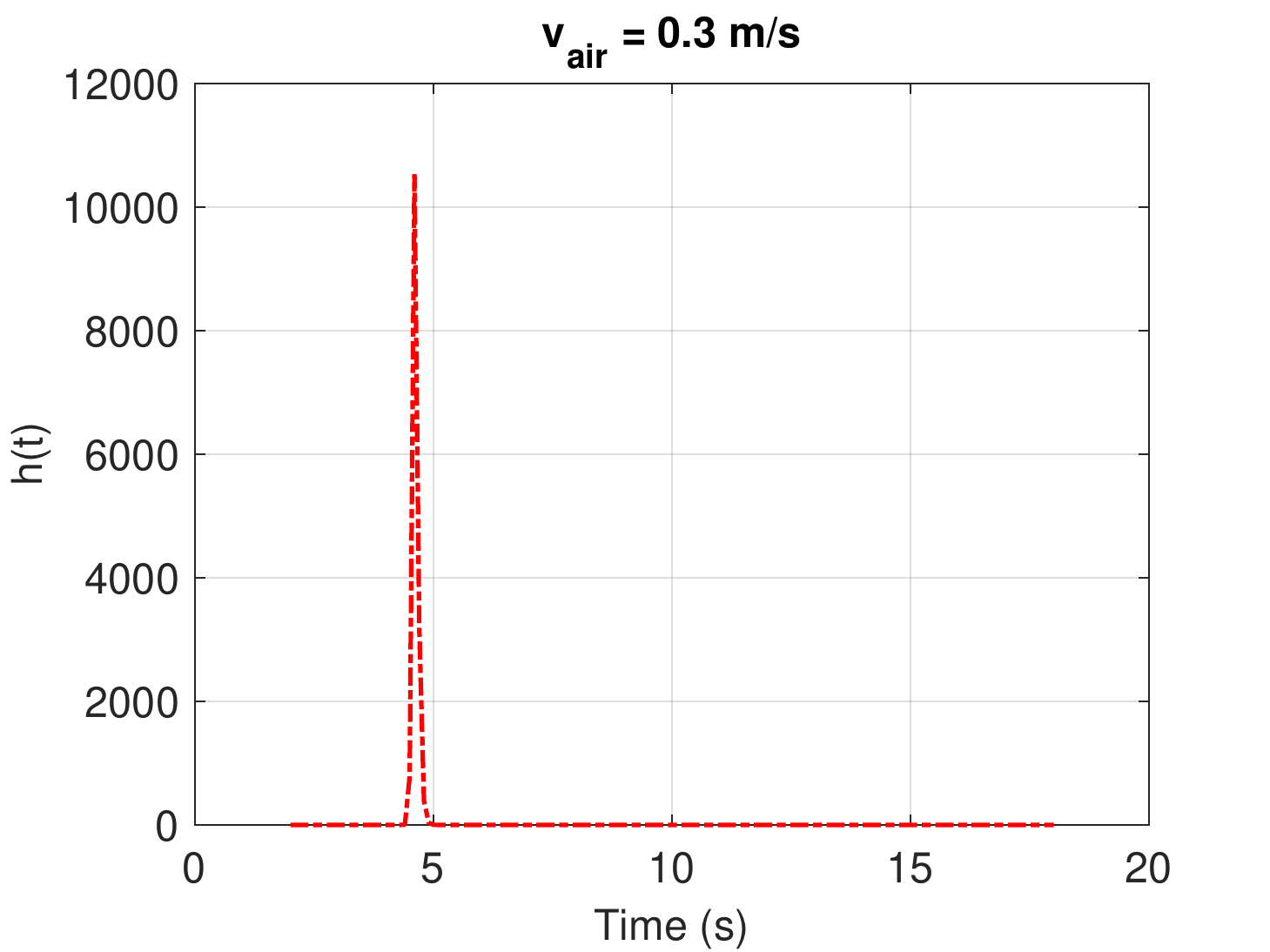}    
	\includegraphics[width=0.325\textwidth]{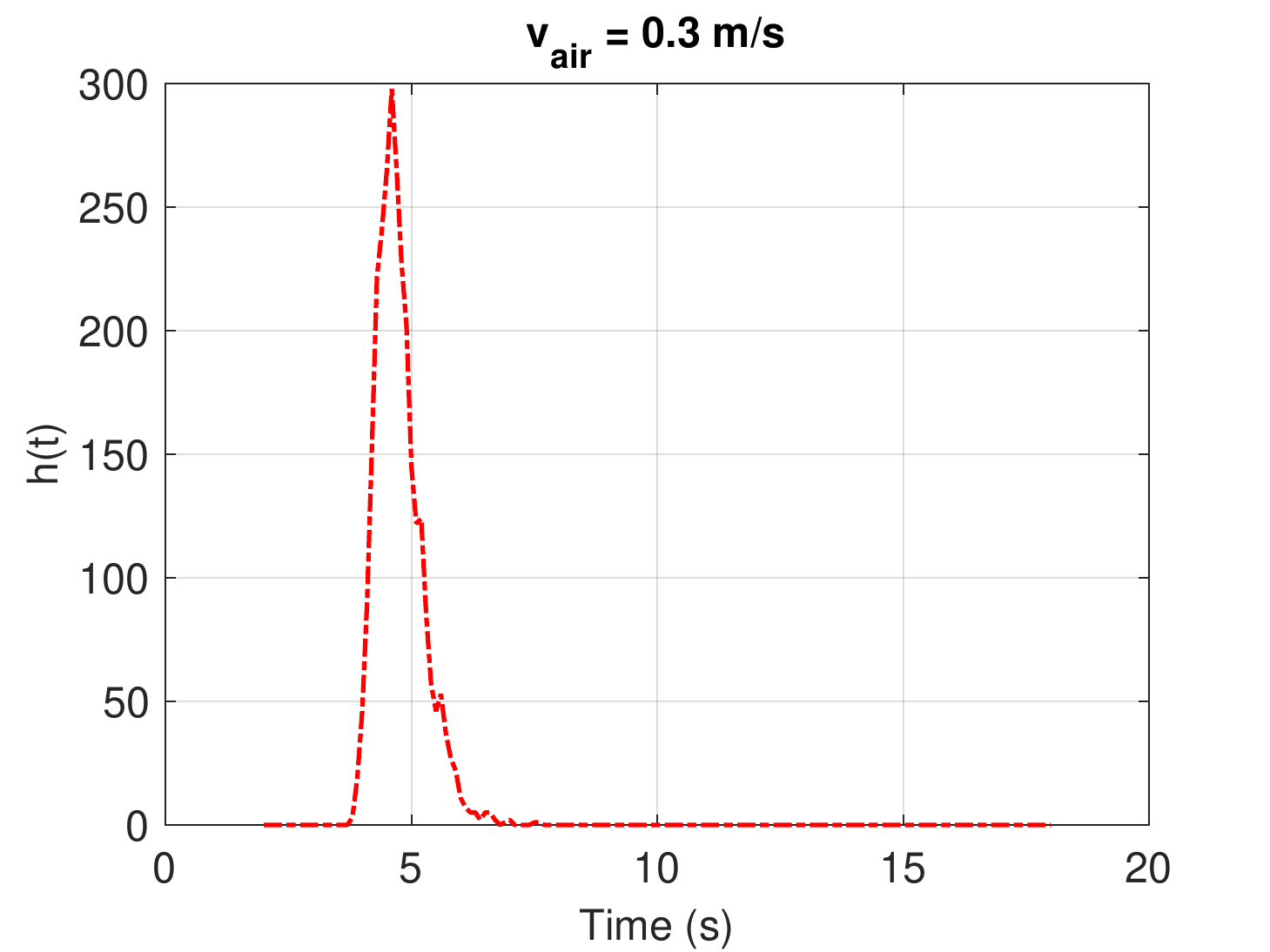}  
	\includegraphics[width=0.325\textwidth]{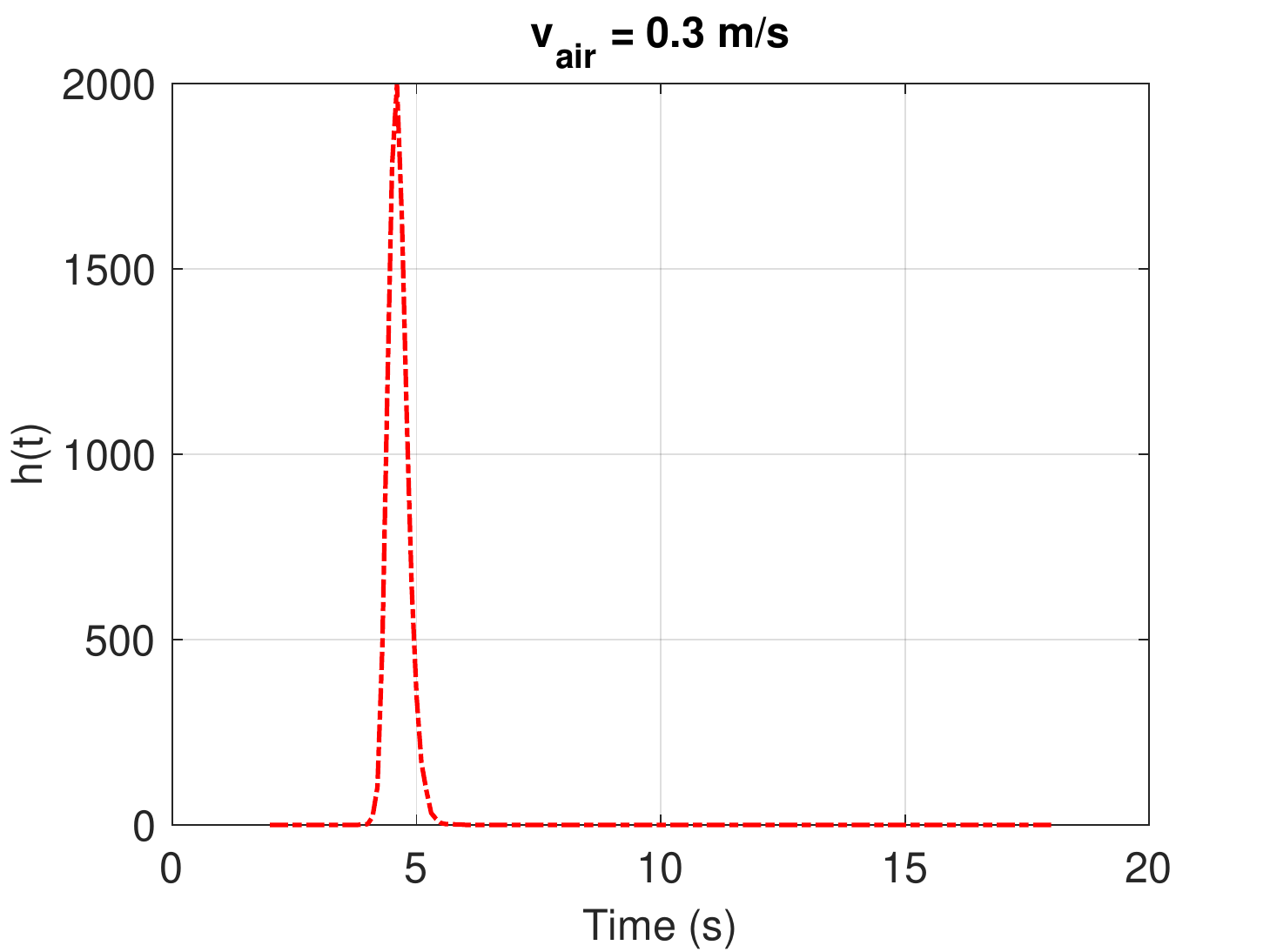} \\
	\scriptsize \hspace{0.1 in}  (a) \hspace{2.2 in}  (b) \hspace{2.2 in} (c) \\
	\includegraphics[width=0.325\textwidth]{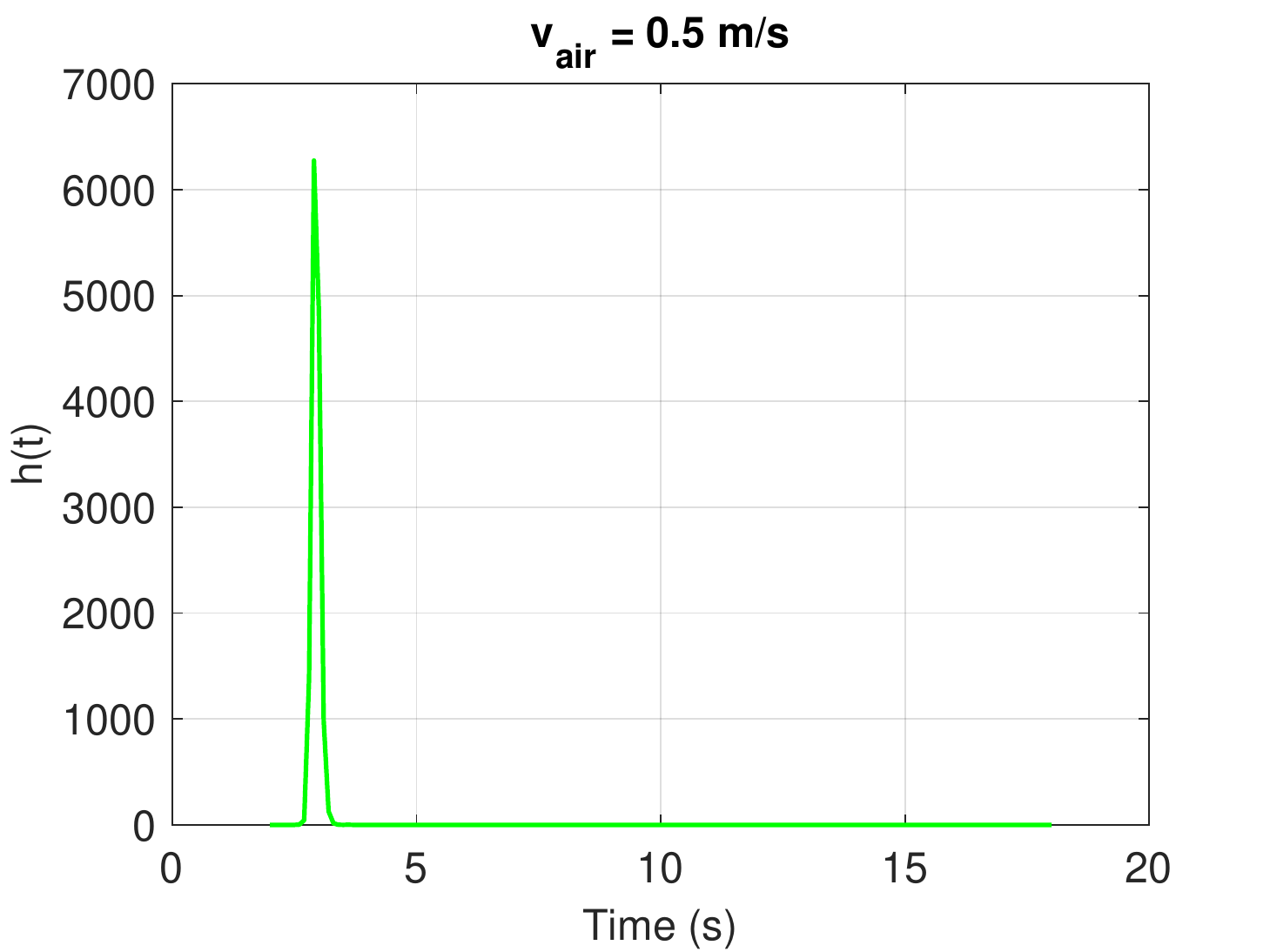}    
	\includegraphics[width=0.325\textwidth]{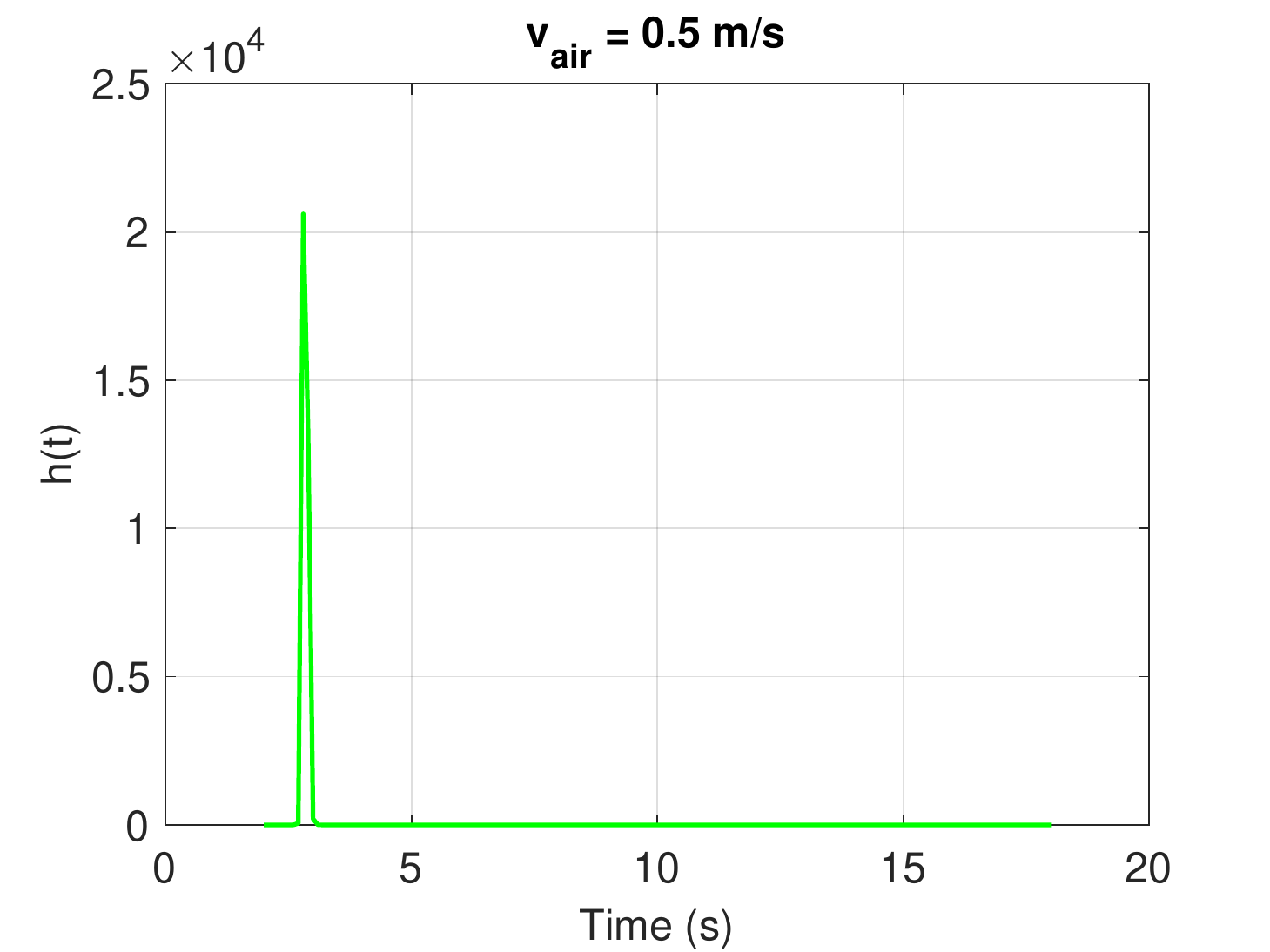}  
	\includegraphics[width=0.325\textwidth]{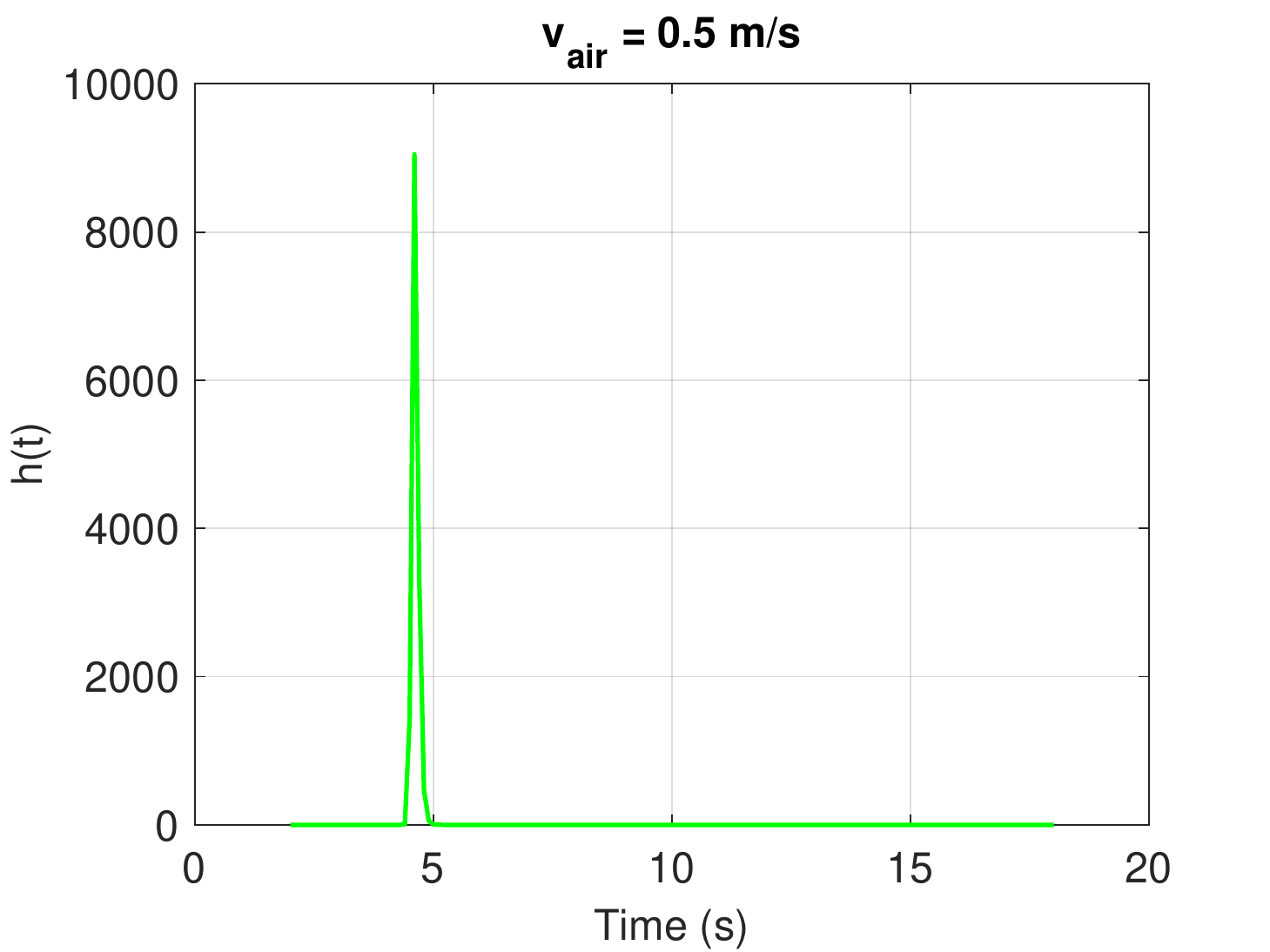} \\
	\scriptsize \hspace{0.1 in}  (a) \hspace{2.2 in}  (b) \hspace{2.2 in} (c) \\
	\caption{Samples of the end-to-end impulse response for the turbulent MC channel when (a-c) $v_{air} = 0.1$ m/s (d-f) $v_{air} = 0.3$ m/s (g-i) $v_{air} = 0.5$ m/s.}
	\label{EIR}
\end{figure*}

\subsection{Analysis}
Although the multi-modal distributions in Fig. \ref{pdf123} may seem counter-intuitive at the first glance, they actually reflect the very complicated nature of airborne transmission due to the turbulent flows with different particle sizes. Firstly, larger particles have a ballistic trajectory due to gravity as also observed in Fig. \ref{Flow}. Secondly, the smaller particles tend to suspend in the air, and they are entrained by the constant and turbulent airflows. Since the characteristic of turbulent flows is inherently stochastic, every emission with the same parameters can lead to different trajectories in the channel. In \cite{trivedi2021estimates}, an analysis that investigates the stochasticity of the particle dispersion due to a turbulent cough emission shows that the spatial probability distributions of the emitted particles have multi-modal distributions in the air. Furthermore, the results and analysis in \cite{gulec2020distance} shows that there can be unexpected particle velocities in the MC channel due to the turbulence in the vicinity of the TX. Therefore, these effects can lead to multi-modal distributions in the reception.

In Fig. \ref{Dia}, the probabilities of particle reception with respect to their diameters are given. In parallel with the observations in Figs. \ref{Flow} and \ref{Reception}, Fig. \ref{Dia} shows that mostly aerosols arrive at the RX, which is based on statistical data obtained by CFD simulations. Although there are some large received particles up to $25$ $\mu$m for $v_{air} = 0.1$ m/s and $v_{air} = 0.3$ m/s and $41$ $\mu$m for $v_{air} = 0.5$ m/s, their reception probabilities are negligible and most of the received particles are smaller than $6$ $\mu$m. Moreover, while the reception probability decreases as the particle diameter increases as given in Figs. \ref{Dia} (a) and (b), this is not the case for $v_{air} = 0.5$ m/s as shown with the chaotic distribution of diameters in Fig. \ref{Dia} (c). This chaotic distribution of diameters depicts the fact that the effect of turbulence still continues due to the increased velocity of particles by the ambient airflow. 

The resulting plot of $ P_{inf} $ values given in (\ref{P_inf_eq1})-(\ref{P_inf_eq3}) for different $\gamma$ values are shown in Fig. \ref{P_inf}. These results show that larger air velocities increase the infection probability for susceptible people for a face-to-face scenario. Even though $\gamma$, i.e., the immune system's strength of a susceptible human, is high, the risk of infection still continues due to the higher exposure of pathogen-laden particles. The results in Fig. \ref{P_inf} also show the effect of aerosols which are not taken into account before in the literature for the calculation of the probability of infection in addition to the effect of large droplets in smaller numbers. This derivation of $P_{inf}$ shows that modeling the reception of particles is non-trivial due to the highly complex nature of turbulence and it shows that multi-modal distributions lie under this turbulent complexity. These analytical derivations also can be employed in epidemiology studies with more realistic parameters for interhuman airborne pathogen transmission. 

For real-life scenarios, airborne pathogen transmission can be more complicated, since the infection actually includes the interactions of pathogen-laden particles at the reception regions into the human body such as mouth, nose, and eyes. Furthermore, pathogens interact with the immune system and the vaccinations can be also effective in the process ending with infection. These are open research issues which are also covered within the broader mobile human ad hoc network framework given in \cite{gulec2022mobile}. In this paper, our focus is to investigate the effect of turbulent flows on the reception. Therefore, we employ a simple reception model. As for the channel, the Navier-Stokes equations are accepted as the standard method to model the turbulent flows, although it is known that it cannot capture the nature of turbulent flows in every scenario \cite{cebeci2005computational}. In the next section, we focus on the characterization of the turbulent MC channel by using obtained data in the CFD simulations.

\section{End-to-End Impulse Response Characterization for the Turbulent MC Channel} \label{EtEIRC}
In this section, we focus on the characterization of the turbulent MC channel between the TX and RX. To this end, the obtained data via CFD simulations are post-processed and the end-to-end impulse responses are generated. Then, the pdfs for different air velocities are estimated as done in Section \ref{PoI}.

\begin{figure*}[tb]
	\centering
	\includegraphics[width=0.325\textwidth]{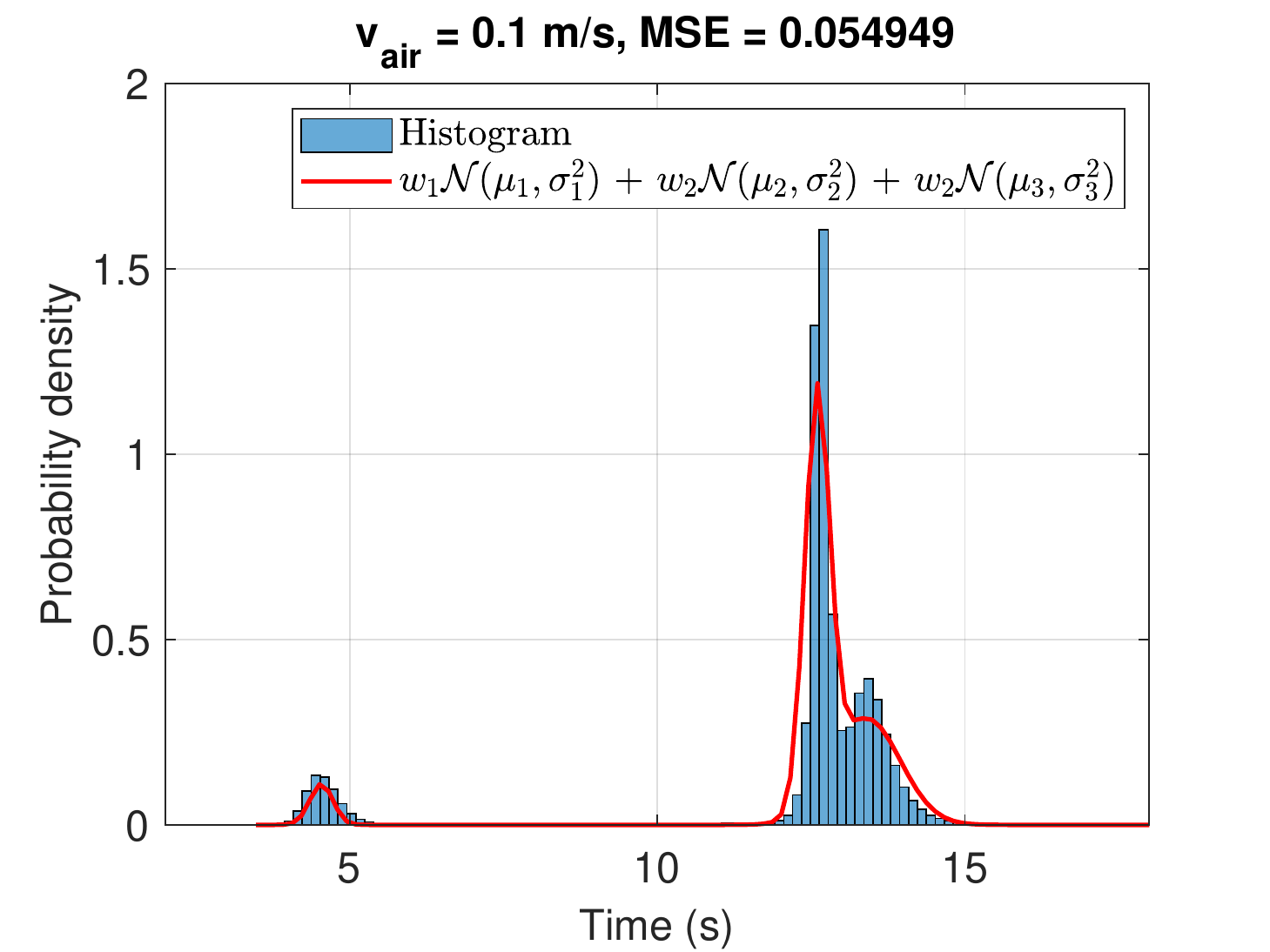}    
	\includegraphics[width=0.325\textwidth]{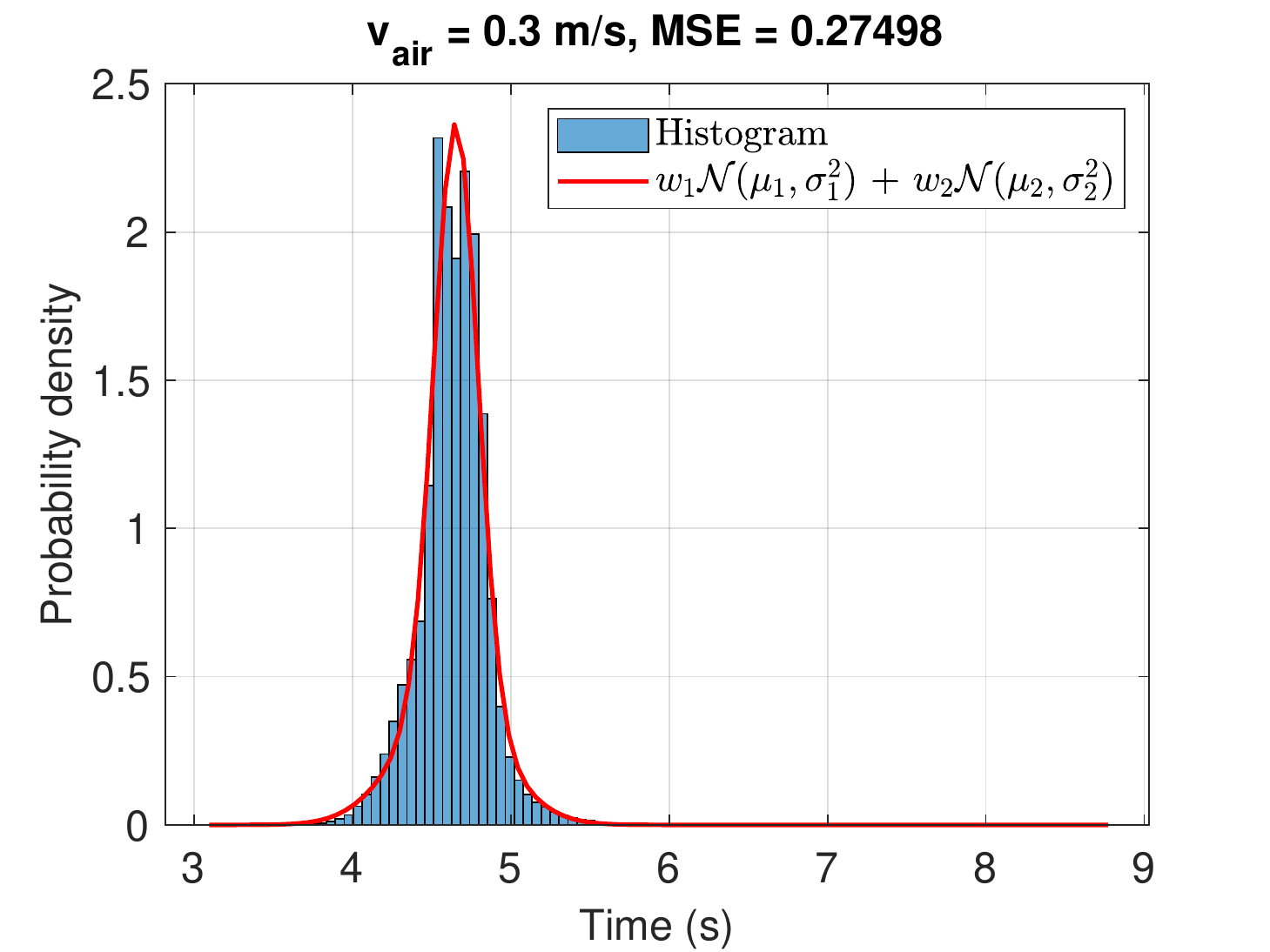}  
	\includegraphics[width=0.325\textwidth]{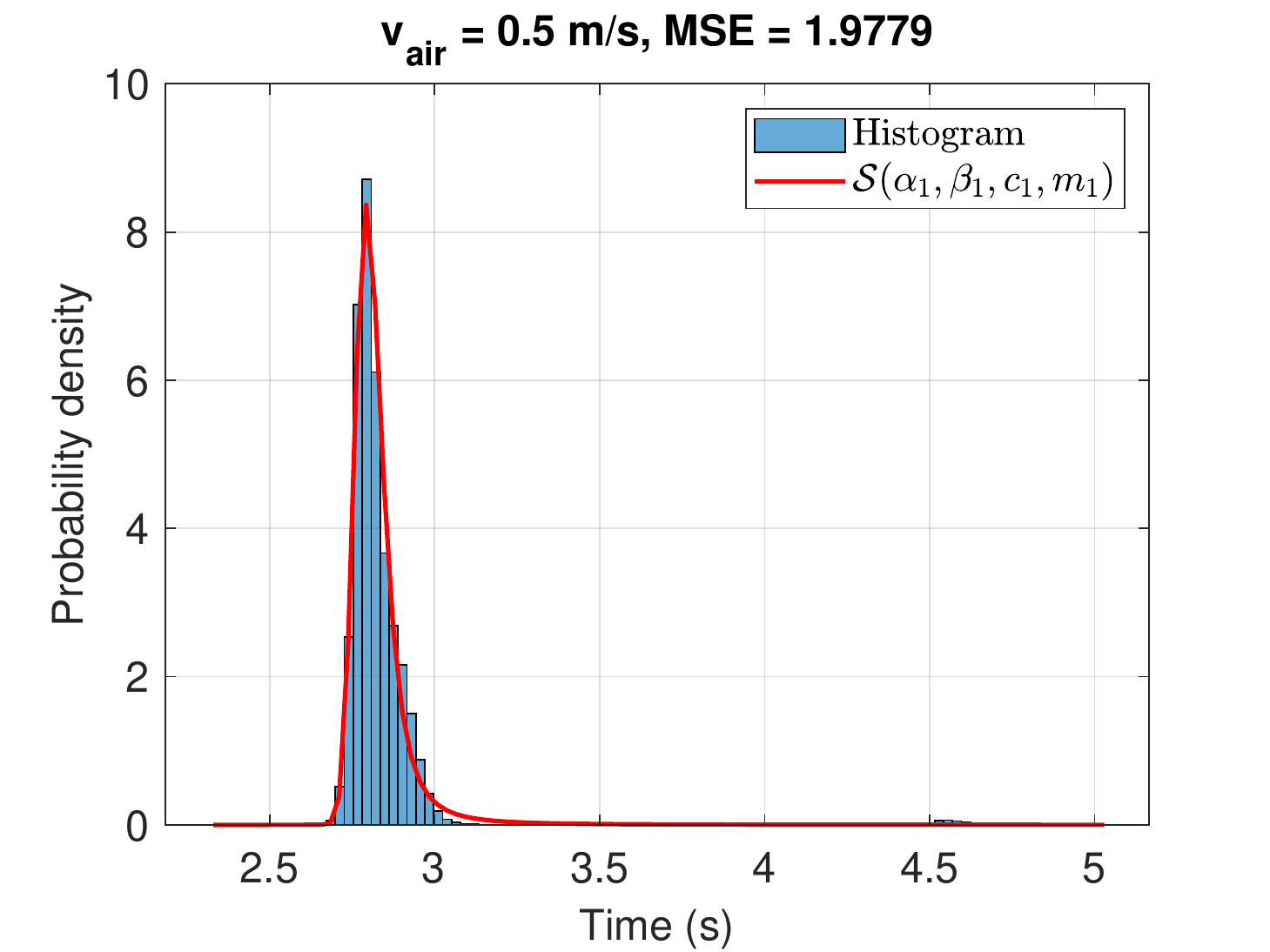} \\
	\scriptsize \hspace{0.1 in}  (a) \hspace{2.2 in}  (b) \hspace{2.2 in} (c) \\
	\caption{Probability density function of $ h(t) $ for (a) $v_{air} = 0.1$ m/s (b) $v_{air} = 0.3$ m/s (c) $v_{air} = 0.5$ m/s.}
	\label{EIRpdf}
\end{figure*}

\begin{table*}[bt]
	\centering
	\caption{Estimated statistical parameters of 	$ f_{h_{i}} $}
	\scalebox{1}{
		\begin{tabular}{p{15pt}|p{10pt}|p{40pt}|p{40pt}|p{10pt}|p{40pt}|p{40pt}|p{10pt}|p{40pt}|p{40pt}|p{40pt}} \hline
			Pdf	& $ w_1 $ & \multicolumn{2}{|l|}{Distribution 1}                                   & $ w_2 $ & \multicolumn{2}{|l|}{Distribution 2}                                  & $ w_3 $ & \multicolumn{2}{|l|}{Distribution 3}                & MSE      \\ \hline
			$ f_{h_{1}} $ & $ 0.05 $ & $\mu_1$=$4.54$ & $\sigma_1$=$0.19$ & $ 0.55 $ & $\mu_2$=$12.6 $  & $ \sigma_2$=$0.2 $   & $ 0.4 $    & $\mu_3$=$13.4$           & $\sigma_3$=$0.55$                                   & $ 0.054949 $ \\ \hline
			$ f_{h_{2}} $ & $ 0.68 $ & $\mu_1$=$4.65$ & $\sigma_1$=$0.14$ & $ 0.32 $ &$\mu_2$=$4.61 $  & $ \sigma_2$=$0.32 $ & - & - & - & $ 0.27498 $ \\ \hline
			$ f_{h_{3}} $ & - & $\alpha_1$=$1.47$, $\beta_1$=$0.99$ & $c_1$=$0.03 $, $ m_1$=$2.8 $ & -  & - & - & -    & -           & -                                   & $ 1.9779 $ \\ \hline
	\end{tabular}}
	\label{EIR_Est_parameters}
\end{table*}

Since the particles from the TX are emitted in a very short emission time, the input to the turbulent MC channel can be considered as an impulse-like emission. Therefore, the received signal, which is the number of received particles at the RX, can be considered as the end-to-end impulse response of the MC system ($h(t)$) as shown in Fig. \ref{EtE} \cite{gulec2020droplet}. In Fig. \ref{EIR}, nine different end-to-end impulse responses where the rows correspond to the air velocities in the MC channel as $0.1$ m/s (a-c), $0.3$ m/s (d-f) and $0.5$ m/s (g-i) are shown. In this figure, each row depicts three different samples of $h(t)$ among $500$ trials for the same air velocity in the MC channel. In Fig. \ref{EIR} (a)-(c) for $v_{air} = 0.1$ m/s, it is observed that the system response can be in very different signal shapes. While $h(t)$ has an approximately symmetric shape like a Gaussian function in Fig. \ref{EIR} (a), it can have a more skewed shape as a long-tailed function as shown in Fig. \ref{EIR} (c). In addition, it can also have a longer response time ($\approx$10 s) and relatively much lower amplitude than other responses as given in Fig. \ref{EIR} (b). Moreover, the peak arrival times of the particles show a large variation in Figs. \ref{EIR} (a)-(c). The results for $v_{air} = 0.1$ m/s are important to understand the effect of turbulence caused by the emission in an indoor environment without ventilation, since there is nearly always a slight airflow in air-based indoor MC channels. As $v_{air}$ increases in the turbulent MC channel, $h(t)$ resembles to a more impulse-like function. Nevertheless, the variation of the reception times are large due to the turbulence.

Based on the results for different air velocities, the histograms and the estimated pdfs are shown in Fig. \ref{EIRpdf}. Similar to the pdfs of received number of particles in Fig. \ref{pdf123}, the estimated pdfs are characterized by weighted multi-modal distributions except for $v_{air} = 0.5$ m/s. Based on the visual inspection of the histograms, it is decided which distribution to use for the pdf estimation. These weighted pdfs are given by
\begin{align}
	\hspace{-0.2cm}	f_{{h_1}}(t) \hspace{-0.05cm} &= \hspace{-0.05cm} w_1 \mathcal{N}(\mu_1,\sigma_1^2) \hspace{-0.1cm} + \hspace{-0.1cm} w_2 \mathcal{N}(\mu_2,\sigma_2^2) + \hspace{-0.1cm} w_3 \mathcal{N}(\mu_3,\sigma_3^2)\label{EIR_pdf_eq1}\\
	\hspace{-0.2cm}	f_{{h_2}}(t) \hspace{-0.05cm} &= \hspace{-0.05cm} w_1 \mathcal{N}(\mu_1,\sigma_1^2) \hspace{-0.1cm} + \hspace{-0.1cm} w_2 \mathcal{N}(\mu_2,\sigma_2^2) \hspace{-0.1cm}  \label{EIR_pdf_eq2} \\
	\hspace{-0.2cm}	f_{{h_3}}(t) \hspace{-0.05cm} &= \mathcal{S}(\alpha_1, \beta_1, c_1, m_1),
	\label{EIR_pdf_eq3}
\end{align}
where $ f_{{h_1}}(t) $, $ f_{{h_2}}(t) $, $ f_{{h_3}}(t) $ are the pdfs for $v_{air} = 0.1$ m/s, $v_{air} = 0.3$ m/s, and $v_{air} = 0.5$ m/s, respectively, $w_1$, $w_2$, $w_3$ are weight coefficients, $\mathcal{N}(\mu,\sigma^2)$ show a normal distribution with the mean $\mu$ and standard deviation $\sigma$, and  $ \mathcal{S}(\alpha_1, \beta_1, c_1, m_1) $ shows a stable distribution where $\alpha_1$ is the stability parameter, $\beta_1$ is the skewness parameter, $c_1$ is the scale parameter, and $m_1$ is the location parameter. The pdf of the stable distribution is not analytically expressible in general. Instead, this pdf is defined via the inverse Fourier transform of its characteristic function ($\Phi(x)$) as given by \cite{nolan2020univariate}
\begin{equation}
	f_{{h_3}}(t) = \frac{1}{2\pi} \int_{-\infty}^{\infty} \Phi(x) \textrm{e}^{-jxt} dx.
\end{equation}
Here the characteristic function is analytically given as
\begin{equation}
	\Phi(x) = \textrm{exp} \left(j x m_1-|c_1 x|^{\alpha_1}(1 - j \beta_1 \textrm{sgn}(x) \kappa)\right),
\end{equation}
where $ \textrm{sgn}(x) $ is the sign function and $\kappa$ is defined by
\begin{subequations}
	\begin{numcases}
		{\kappa =}
		\textrm{tan} \left(\frac{\pi \alpha_1}{2} \right), & $ \alpha_1 \neq 1$ \\
		- \frac{2}{\pi} \textrm{log}|t|, & $ \alpha_1 = 1$.
	\end{numcases}
\end{subequations}

For the pdfs given in (\ref{EIR_pdf_eq1})-(\ref{EIR_pdf_eq3}), the parameters are estimated with the maximum likelihood estimation method as explained in Section \ref{Particle_Distr}. These estimated parameters are given with their MSE in Table \ref{EIR_Est_parameters}.

As shown in Fig. \ref{EIRpdf}, the end-to-end system responses are not characterized similar to a diffusion MC channel with drift where the flow velocity changes the system response with the same analytical expression as reviewed in \cite{jamali2019channel}. The obvious distinction of turbulent air-based MC channel from the diffusion MC channels with drift is the high initial velocity of particles, gravity and the varying sizes of particles. The high initial velocity causes the turbulent flows and the corresponding air-particle interactions. Furthermore, the gravity has a similar role to a filter in the MC channel by eliminating large particles by settling to the ground. As for the small particles, i.e., aerosols, they suspend in the air in the abscence of an airflow. In our case, they are entrained by the turbulent airflows. As given in Fig. \ref{Dia}, when $v_{air}$ is smaller, the received particles are dominated by smaller particles. As $v_{air}$ increases, the reception is dominated by larger particles with a shorter arrival time at the RX. Hence, the arrival times for $v_{air}=0.5$ m/s can be explained with a uni-modal distribution. However, when the ambient air velocity is not strong enough, the particle cloud is more dispersed and the effect of turbulence in the vicinity of the TX can be amplified as particles move closer to the RX. Hence, it causes large variations in the arrival times ending up with multi-modal distributions.

The results for the characterization of turbulent flows in a MC channel can be employed for the further analysis and implement new techniques in different turbulent MC scenarios. For instance, an information theoretical analysis can be useful for the analysis of air-based turbulent MC channels as applied in \cite{hoeher2021mutual, abbaszadeh2021kolmogorov}. Furthermore, the encoding of information in turbulent MC scenarios can be developed with the obtained results in this paper.

\section{Conclusion} \label{Conc}
In this paper, the airborne transmission of pathogens emitted via coughing between two humans is modeled with a CFD approach incorporating turbulent airflows and turbulent dispersion of droplets and aerosols. By using the statistical data collected with CFD simulations, it is revealed that turbulence cause multi-modal distributions for the pdf of received particles by the susceptible human. Furthermore, numerical results show that augmented air velocity causes the increment of the probability of infection. Derived probability of infection expressions can be employed in epidemiological models to realistically consider the indoor airborne transmission. On the other hand, the end-to-end system response of the  air-based turbulent MC system is characterized. It is shown that the system response also shows multi-modal distributions except for the larger ambient air velocity. The characterization of the turbulent MC channel can also be used for the realistic design of MC systems with turbulent flows. As the future work, it is planned to extend this study by including scenarios with masks and breathing.


\bibliographystyle{ieeetran}
\bibliography{IEEEabrv,ref_fg_t_apt}

\begin{IEEEbiography}[{\includegraphics[width=1in,height=1.25in,clip,keepaspectratio]{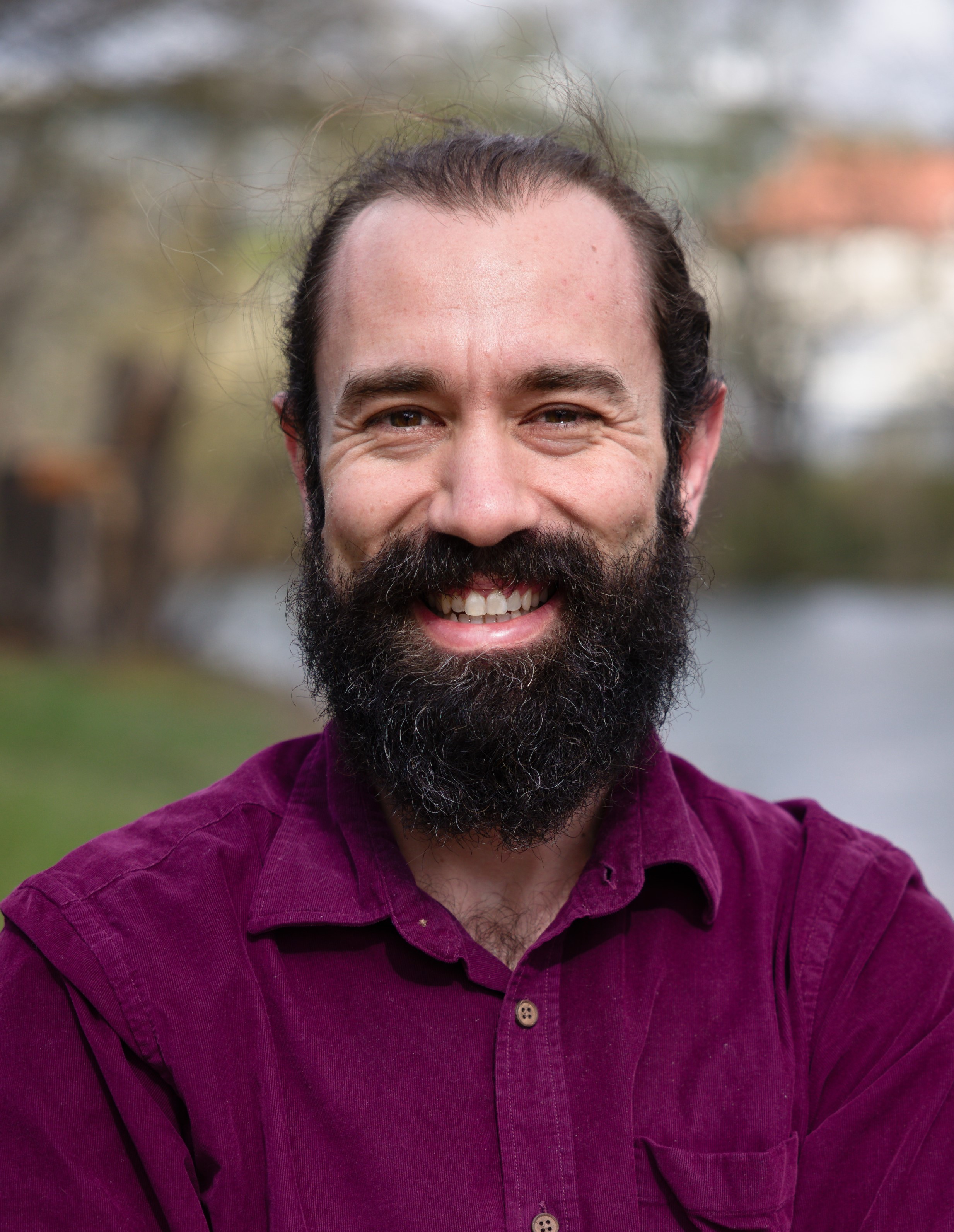}}]{Fatih Gulec} (Member, IEEE) received his B.Sc. and M.Sc. degree from Gazi University, Ankara, Turkey in 2007 and 2015, respectively both in electrical and electronics engineering. He received the Ph.D. degree in İzmir Institute of Technology, İzmir, Turkey in 2021 in electronics and communication engineering. After a research stay as a postdoctoral researcher at the School of Electrical
Engineering and Computer Science, TU Berlin, Germany
with the DAAD scholarship, he is currently with the
Department of Electrical Engineering and Computer
Science, York University, Canada as a postdoctoral research
fellow. His research interests include molecular communications and computational biology.
\end{IEEEbiography}

\begin{IEEEbiography}[{\includegraphics[width=1in,height=1.25in,clip,keepaspectratio]{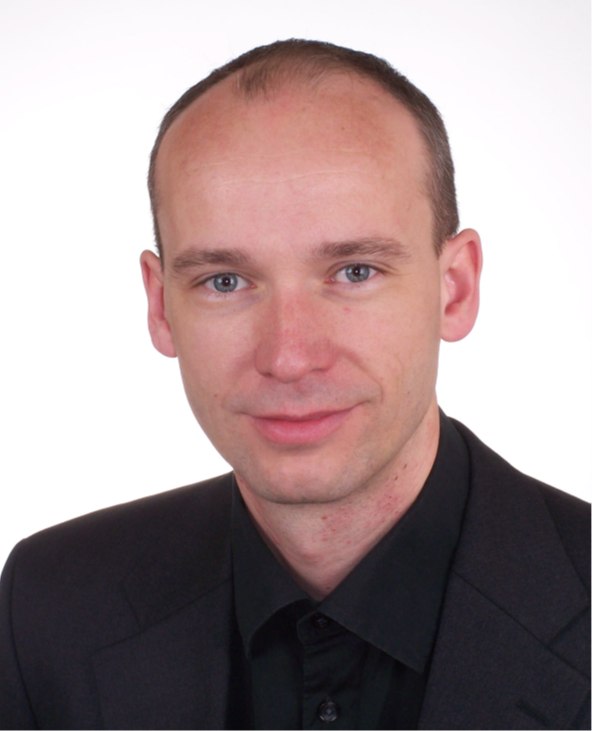}}]{Falko Dressler} (Fellow, IEEE) received his M.Sc. and Ph.D. degrees from the Dept. of Computer Science, University of Erlangen in 1998 and 2003, respectively. He is a full professor and Chair for Data Communications and Networking at the School of Electrical Engineering and Computer Science, TU Berlin. Dr. Dressler has been associate editor-in-chief for IEEE Trans. on Mobile Computing and Elsevier Computer Communications as well as an editor for journals such as IEEE/ACM Trans. on Networking, IEEE Trans. on Network Science and Engineering, Elsevier Ad Hoc Networks, and Elsevier Nano Communication Networks. He has been chairing conferences such as IEEE INFOCOM, ACM MobiSys, ACM MobiHoc, IEEE VNC, IEEE GLOBECOM. He authored the textbooks Self-Organization in Sensor and Actor Networks published by Wiley \& Sons and Vehicular Networking published by Cambridge University Press. He has been an IEEE Distinguished Lecturer as well as an ACM Distinguished Speaker. Dr. Dressler is an IEEE Fellow as well as an ACM Distinguished Member. He is a member of the German National Academy of Science and Engineering (acatech). He has been serving on the IEEE COMSOC Conference Council and the ACM SIGMOBILE Executive Committee. His research objectives include adaptive wireless networking (radio, visible light, molecular communications) and embedded system design (from microcontroller to Linux kernel) with applications in ad hoc and sensor networks, the Internet of Things, and cooperative autonomous driving systems.
\end{IEEEbiography}

\begin{IEEEbiography}[{\includegraphics[width=1in,height=1.25in,clip]{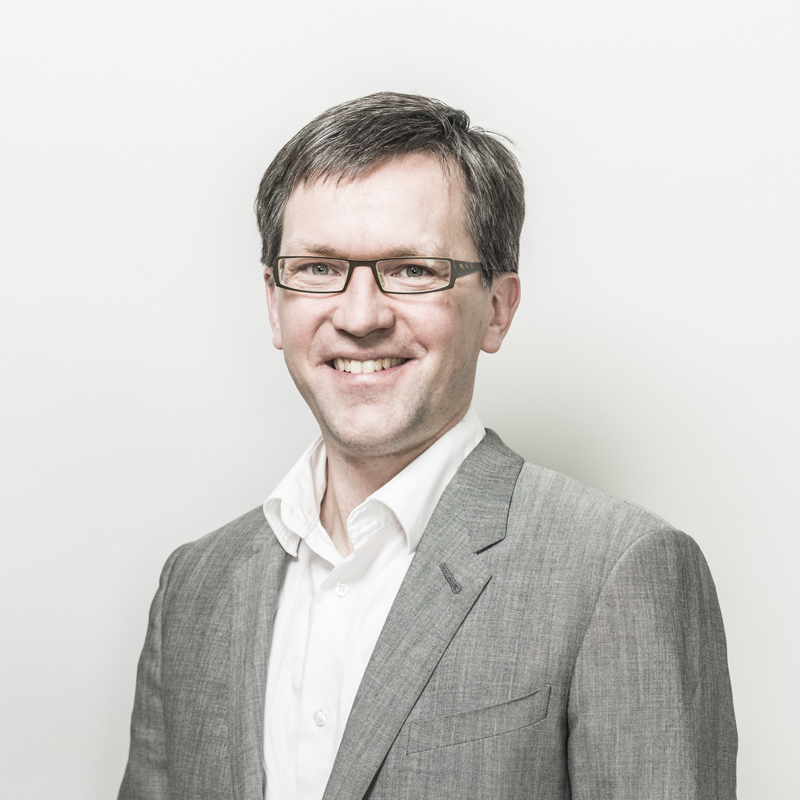}}]{Andrew W. Eckford} (Senior Member, IEEE) received the B.Eng. degree in electrical engineering from the Royal Military College of Canada in 1996, and the M.A.Sc. and Ph.D. degrees in electrical engineering from the University of Toronto in 1999 and 2004, respectively. He was a Postdoctoral Fellowship with the University of Notre Dame and the University of Toronto, prior to taking up a faculty position with York, in 2006. He is an Associate Professor with the Department of Electrical Engineering and Computer Science, York University, Toronto, ON, Canada. He has held courtesy appointments with the University of Toronto and Case Western Reserve University. In 2018, he was named a Senior Fellow of Massey College, Toronto. He is also a coauthor of the textbook Molecular Communication (Cambridge University Press). His research interests include the application of information theory to biology and the design of communication systems using molecular and biological techniques. His research has been covered in media, including The Economist, The Wall Street Journal, and IEEE Spectrum. His research received the 2015 IET Communications Innovation Award, and was a Finalist for the 2014 Bell Labs Prize.
\end{IEEEbiography}

\end{document}